\documentclass[12pt,aps,amsmath,latexsym,amsfonts]{JHEP3}

\usepackage{epsfig}
\usepackage{amsfonts,amssymb,amsmath}

\newcommand{\be}{\begin{equation}}
\newcommand{\ee}{\end{equation}}
\newcommand{\bea}{\begin{eqnarray}}
\newcommand{\eea}{\end{eqnarray}}

\def\E{\text{\,\texttt{E}\,}}
\def\T{\text{\,\texttt{T}\,}}

\preprint{DCPT-08/25}

\title{Braneworld Black Holes}
\author{Ruth Gregory\\
Centre for Particle Theory, 
Durham University, South Road, Durham, DH1 3LE, UK.}
\abstract{
In these lectures, I give an introduction to and overview of braneworlds
and black holes in the context of warped compactifications. I first
describe the general paradigm of braneworlds, and introduce the Randall-Sundrum
model. I discuss braneworld gravity, both using perturbation theory, and 
also non perturbative results. I then discuss black holes on the brane, 
the obstructions to finding exact solutions, and ways of tackling these
difficulties. I describe some known solutions, and conclude with some
open questions and controversies.}

\begin{document}
\section{Introduction} \label{intro}

Nearly a century ago, Kaluza and Klein theorized that by adding an
extra dimension to space, you could unify electromagnetism with gravity.
Thus our first `unified theory' was born -- at the price of an
extra unseen dimension. Nowadays, extra dimensions are 
an integral part of fundamental
theoretical physics, and the consequences of devising consistent 
means of hiding these extra dimension has led to an explosion of 
activity in recent years in string theory, cosmology, and phenomenology.
Braneworlds are just part of this general story, and represent a
particular way of dealing with the extra dimensions that is empirical,
but precise and calculable. They have proved indispensable for 
developing ideas and methods which have then been used in more
esoteric but fundamentally grounded models in string theory. These lectures
are about braneworlds, and deal with the deeply interesting, but thorny
issue of how to describe braneworld black holes. 

Simply put, a braneworld is a slice through spacetime on which we live. 
We cannot (easily) see the extra dimensions perpendicular to our slice, 
as all of our standard physics is confined. 
We can however, deduce those extra dimensions by carefully monitoring 
the behaviour of gravity.
Confinement to a brane may at first sound counter-intuitive,
however, it is in fact a common occurrence. The first braneworld scenarios
\cite{EBW} used topological defects to model the braneworld, with
condensates and zero-modes producing confinement. In 
string theory, D-branes have `confined' gauge theories on their
worldvolumes \cite{DBI} and heterotic M-theory has a natural domain wall
structure \cite{HW}. 

The new phenomenology of braneworld scenarios is primarily located
in the gravitational sector, with a particularly nice geometric
resolution of the hierarchy problem \cite{ADD}.
The scenario has however far outgrown these
initial particle phenomenology
motivations, and has proved a fertile testbed for new possibilities
in cosmology, astrophysics, and quantum gravity.
One of the most popular models with warped extra dimensions is that of
Randall and Sundrum (RS), \cite{RS}, which
consists of a domain wall universe living in five-dimensional
anti-de Sitter (adS) spacetime, and will be the setting for these
lectures. Interestingly, although the RS model is an empirical 
braneworld set-up it can be related to, or motivated by, string theory
in several ways. First of all, it is notionally similar to the heterotic
M-theory set-up, in that the initial RS model had two walls at the end
of an interval. However, this similarity is notional only, and calculationally,
the gravitational spectrum of GR in five dimensions is very
different from the spectrum of low energy heterotic M-theory \cite{HW}.
A more fruitful and robust parallel occurs with type IIB string
theory, where the RS model can (in some rough sense) be associated with 
the near horizon limit of a stack of D3 branes.  Viewed in this context, 
the RS model provides an excellent opportunity to use and test ideas 
from the gauge/gravity or adS/CFT correspondence \cite{MAL}.

The RS model is however particularly valuable as a concrete and
explicit calculational testbed for any theory with extra dimensions
in which gravity is able to probe and modify these hidden directions.
One of the problems with having extra dimensions is that we have to
hide evidence of their existence. We not only have to reproduce 
gravitational and standard model physics on the requisite scales, but 
we also have to ensure that we do not create any 
additional unwanted physics. With RS, the gravitational physics is 
self-consistent and calculable. We can therefore compute
the cosmological and astrophysical consequences of the extra dimension
in a wide variety of physically interesting cases.

Black holes are perhaps the most interesting physical object to
explore within the braneworld framework of extra dimensions. From
the Kaluza-Klein point of view, extra dimensions show up as extra
charges black holes can carry from the 4D point of view \cite{KKBH},
however, in these solutions the black hole is `smeared' along the 
extra dimension rather than localized. Braneworld scenarios are
the antithesis of KK compactifications, consisting of highly localized
and strongly warped extra dimensions, and therefore the implications of
this strongly localized and gravitating brane for black hole physics
are of particular physical and theoretical interest. We now have
compelling evidence of the existence of black holes in nature, from
stellar sized black holes in binary systems, observed via 
X-ray emission from accretion discs \cite{xray}, to supermassive
black holes at the centre of galaxies \cite{smass}, which in the 
case of our own Milky Way can be seen quite clearly from stellar
orbits \cite{milky}. As observational evidence accrues and becomes more
robust, the bounds on the innermost stable orbit of the black hole
(obtained from iron emission lines \cite{Fe}) may eventually start
to confront the theoretical limit from the 4D Kerr metric, and 
possibly provide signatures of extra dimensions, for which the
bounds can be quite different \cite{GH}.  

Turning to the small scale, and taking seriously the possibility 
that braneworlds can provide a resolution of the hierarchy 
problem via a geometric renormalization of the Newton constant \cite{ADD},
raises the possibility that mini black holes can be produced in particle
collisions \cite{LHCBH}. Understanding the formation and decay of these
highly energetic black holes will then allow us to predict signatures
for black hole formation at the LHC \cite{BHLAB}, and is the topic
of a companion set of lectures at this school \cite{YK}.

Finally, there is also a compelling theoretical reason for studying
braneworld black hole solutions, and that is the parallel between
the RS model and the adS/CFT conjecture \cite{Gubser,RSMal,DL}. 
As we explore more
concretely in section \ref{sec:adscft}, by taking the 
near horizon limit of a stack of D3-branes,
the RS model can be thought of as cutting off the spacetime
outside the D-branes; the adS curvature of the RS bulk
is therefore given rather precisely in terms of the
D3 brane charge and the string scale.  Thus, we might expect a
parallel between classical braneworld gravity, and
quantum corrections on the brane. The possibility of finding
a calculational handle for computing the back reaction of
Hawking radiation \cite{Hawking,back,PHK} is extremely attractive,
and of course can potentially feed back into the issue of mini
black holes at the LHC. 

In these lectures, we will review the current status of black
hole solutions in the Randall-Sundrum model, first reviewing
the framework in some detail, concentrating on gravitational
issues, and the link with adS/CFT and holography. We will see
why it is so difficult to find an exact solution, before covering
approximate methods and solutions for brane black holes. Finally,
we describe objections to the holographic picture and some
recent developments in the closely related Karch-Randall \cite{KR}
set-up.

\section{Some Randall-Sundrum Essentials}
\label{sec:RS}

The Randall-Sundrum model has one (or two) domain walls situated
as minimal submanifolds in adS spacetime. In its usual form,
the spacetime is
\begin{equation}
ds^2 = e^{-2k|z|} \left [ dt^2 - d{\bf x}^2 \right ] - dz^2\,,
\label{rsmet}
\end{equation}
where $k = L^{-1}$ is the inverse curvature radius of the negatively curved
5D adS spacetime.
Here, the spacetime is constructed so that there are four-dimensional
flat slices stacked along the fifth $z$-dimension, which have a
$z$-dependent conformal pre-factor known as the warp factor. Since
this warp factor has a cusp at $z=0$, this indicates the presence of a
domain wall -- the braneworld -- which represents an exactly flat
Minkowski universe. The reason for choosing this particular slicing
of adS spacetime is to have a flat Minkowski metric on the brane,
i.e.\ to choose the ``standard vacuum''.

The RS spacetime is an example of a codimension 1 braneworld, 
where we have one extra dimension. In this case, there is a well
defined prescription for finding gravitational solutions with an
infinitesimally thin brane: the Israel equations \cite{Israel}, which 
are essentially a physicists tool extracted from the Gauss-Codazzi
formalism for the differential geometry of submanifolds.
Since this formalism is so widely used, it is worth reviewing it
briefly here (see also \cite{GG}).

In the Israel prescription, we rewrite our 5D spacetime as a 4D base
space, with coordinates $x^\mu$, plus a normal distance, $z$, 
from the ``wall''.
The 4D coordinates remain constant along geodesics normal to the wall, 
thus giving a 5D coordinate system $\{x^\mu,z\}$. This coordinate system
is valid within the radius of curvature of the wall, and splits the
tangent space naturally into parallel and normal components, and
the metric in general has the form
\be
\label{GNdef}
ds^2 = \gamma_{\mu\nu} (x,z) dx^\mu dx^\nu - dz^2\,.
\ee
Choosing the coordinates in this way results in the nontrivial
content of the geometry being located in the four dimensional
metric $\gamma_{\mu\nu}$, and the fifth metric component
is always unity because $z$ is the proper distance from the brane.
$n^a = \delta^a_z$ is the normal to the brane, and
$\gamma_{\mu\nu}|_{z=0}$ is the {\it intrinsic metric} on the brane. Note that
$\gamma_{\mu\nu}$ lies in the tangent bundle of the brane as a manifold
(i.e.\ is a {\it four}-dimensional tensor), and has a five-dimensional
counterpart, the {\it first fundamental form} which we denote 
\be
{\hat \gamma}_{ab} = g_{ab} +n_a n_b = {\rm diag} ( \gamma_{\mu\nu}, 0)
\,.
\ee
$\hat\gamma_{ab}$ is a 5D tensor, but acts as a projection, wiping out
any components orthogonal to the brane.
$\gamma$ and $\hat\gamma$ contain the same physical information, the
distinction is purely mathematical, however we will keep it for the
purposes of this technical discussion.
This particular coordinate or {\it gauge} choice
is called the Gaussian Normal (GN) gauge and is the spacelike 
equivalent of the ADM synchronous gauge.
\FIGURE[ht]{\epsfig{file=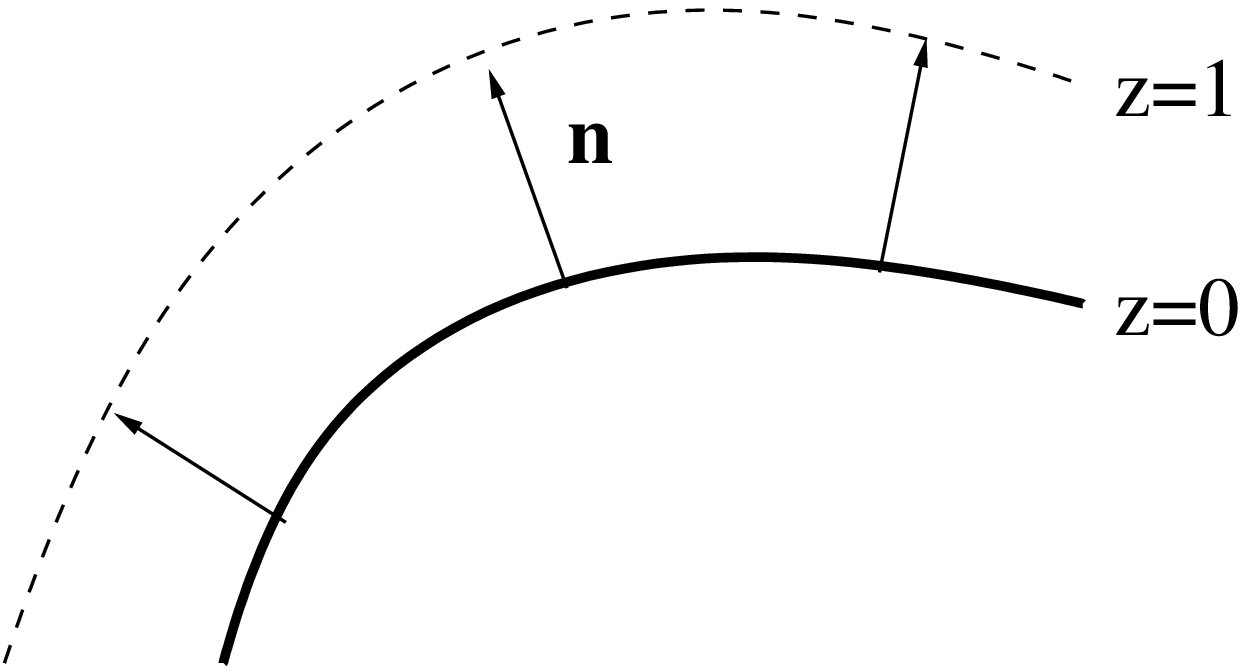,width = 8cm}
\caption{
An illustration of the curved brane and the Gaussian Normal
coordinate system. The brane is the solid line at $z=0$; moving out
a uniform distance from the brane gives a new surface at $z=1$. The
normal to the brane ${\bf n} = \partial / \partial z$ is indicated.
As we move from $z=0$ to $z=1$, distances along the brane will change
in general. This is reflected in the extrinsic curvature (\ref{extcurv}).}
\label{fig:bentbrane}}       % Give a unique label
%\begin{figure}
%\sidecaption[t]
%\includegraphics[scale=.45]{bend.eps}
%\caption{An illustration of the curved brane and the Gaussian Normal
%coordinate system. The brane is the solid line at $z=0$, moving out
%a uniform distance from the brane gives a new surface at $z=1$. The
%normal to the brane ${\bf n} = \partial / \partial z$ is indicated.
%As we move from $z=0$ to $z=1$, distances along the brane will change
%in general. This is reflected in the extrinsic curvature (\ref{extcurv}).}
%\label{fig:bentbrane}       % Give a unique label
%\end{figure}

Surfaces can curve in the ambient manifold, whether or not that is
itself curved (see figure \ref{fig:bentbrane}). This is measured by the 
{\it extrinsic curvature} or {\it second fundamental form}, and is defined via
\be
\label{extcurv}
K_{ab} = {\hat \gamma}^c_a {\hat \gamma}^d_b \nabla_b n_d\,.
\ee
We can use the Riemann identity to get the Gauss-Codazzi relations
\bea
~^{(4)} R^a_{~bcd} &=& {\hat \gamma}^a_e {\hat \gamma}^f_b 
{\hat \gamma}^g_c {\hat \gamma}^i_d
R^e_{~fgi} + K^a_d K_{bc} - K^a_c K_{bd} \\
\Rightarrow
~^{(4)} R _{bd} &=& {\hat \gamma}^a_b {\hat \gamma}^c_d \left ( R_{ac} 
+ R_{aecf} n^e n^f \right ) + K_{ad} K^a_b - K K_{bd}\label{GCRicci}\,.
\eea
In this last relation, we have the 5D Ricci tensor, which
we can replace with the energy momentum tensor via Einstein's equations;
we also have a term which a second use of the Riemann identity allows 
us to write as a Lie derivative of the extrinsic curvature across the brane:
\be
R_{aecf} n^e n^f = - n^b \nabla_b K_{ac} - K^d_c K_{ad}
= - {\cal L}_n K_{ac} + K^d_a K_{cd}\,,
\ee
thus allowing us to rewrite the Gauss-Codazzi equations in terms of
the extrinsic curvature and the energy momentum tensor:
\be
{\cal L}_n K_{ab} = {\hat \gamma}_a^c {\hat \gamma}_b^d T_{cd} 
- {\textstyle{\frac{1}{3}}}
T {\hat \gamma}_{ab} + 2 K^c_a K_{bc} - K K_{ab} - ~^{(4)}R_{ab}\,.
\ee
Therefore, if we imagine our brane to be infinitesimally thin, 
having a distributional energy momentum, $T_{ab}\,\delta (z)$, then
we see that the extrinsic curvature must have a jump across the brane.
Integrating this out, we get the Israel equations:
\be
\label{israeleq}
\Delta K_{ab}  = K_{ab} (z=0^+) - K_{ab} (z=0^-) = 8\pi G_5 \left ( 
T_{ab} - {\textstyle{\frac{1}{3}}} T {\hat \gamma}_{ab} \right )\,.
\ee

Returning to the Randall Sundrum metric, (\ref{rsmet}), we see that 
\be
K_{\mu\nu} = - \Gamma^z_{\mu\nu} n_z = \mp k e^{-2k|z|} \eta_{\mu\nu}\,,
\ee
(where we are now dropping the distinction between the brane tangent
space and the bulk tangent space, as the situation is physically clear).
Using (\ref{israeleq}) we see that the brane has an energy-momentum
tensor proportional to the metric on the brane:
\be
T_{\mu\nu} = \E_{RS} \eta_{\mu\nu} = \frac{6k}{8\pi G_5} \eta_{\mu\nu}\,.
\label{critRStens}
\ee
Notice the very precise form of this energy momentum. First, because it
is proportional to the intrinsic metric, this means that the brane has
tension (rather than pressure) and this tension is exactly equal to
its energy, $\E=\T$. Thus the brane energy momentum has exactly the 
same form as a cosmological constant term on the brane. Second, the
actual value of this tension is precisely related to the bulk cosmological
constant:
\be
\Lambda = -6k^2 = - \frac{(8\pi G_5 \E_{RS})^2}{6}\,.
\label{RSTdef}
\ee
This is sometimes referred to as the fine-tuned, or {\it critical} RS
brane. As we will see later, this relation can be relaxed, leading to
de Sitter or anti de Sitter RS branes (the latter of which are also
known as Karch-Randall (KR) branes \cite {KR}).

\section{Gravity in the Randall Sundrum model}
\label{sec:RSgrav}

Obviously the RS model can only describe our real universe if it
correctly reproduces gravitational physics at experimentally tested
scales. This means we have to be able to reproduce Einstein gravity in
our solar system, and the standard cosmological model for our 
universe. While the Israel equations give us the general formalism
for getting our braneworld metric, finding actual solutions can
be a far trickier matter, as indeed finding a general solution
of Einstein's equations is a tricky matter! We therefore resort,
as with standard gravity, to two main approaches:

$\bullet$ Local Physics, or perturbation theory, and

$\bullet$ ``Big Picture'' or geometry, finding exact solutions
assuming symmetries.

\noindent In either case, we have to accept that gravity on the brane
is a projection of the full higher dimensional
nature of gravity, and is therefore a {\it derived quantity}.

\subsection{Perturbation theory}

In General Relativity (GR), classical perturbation theory 
involves perturbing the metric
\be
g_{ab} \to g_{ab} + h_{ab}
\ee
around a given background solution. There are 3 main issues to bear in mind.

\begin{enumerate}

\item
$h_{ab}$ is a perturbation and should therefore be ``small''. What
does this mean? In practise we have to be careful about our coordinate
system, and always look at $h$ in a regular system. For the Schwarzschild
solution for example, this means using Kruskal coordinates.

\item Gauge freedom: GR has a large gauge group -- physics is invariant
under general coordinate transformations (GCT's),
and there are many gauge degrees of freedom in $h_{ab}$. For example, in 4D,
$h_{ab}$ has 10 independent components, but the graviton has
only 2 physical degrees of freedom.
Under a GCT
\be
X^a \to X^a + \xi^a \; , \;\;\; g_{ab} \to g_{ab} + {\cal L}_\xi g_{ab}\,,
\ee
hence
\be
\delta g_{ab} = \xi_{a;b} + \xi_{b;a}
\ee
and we can use this to make a choice of gauge. A common choice for
relativists is the harmonic gauge
\be
{\bar h}^a_{b;a} = h^a_{b;a} - {\textstyle{\frac{1}{2}}} h^a_{a;b} = 0\,,
\ee
and in vacuo we can also choose $h^a_a=0$: the transverse tracefree (TTF) 
gauge.  Note that this
does not uniquely specify the gauge, e.g.\ $\xi^a_{;a} = 0 =
\nabla^2 \xi^a$ gives an allowed gauge transformation.

\item Finally, we need the perturbation of the Ricci tensor:
\be
\delta R_{ab} = -{\textstyle{\frac{1}{2}}} \nabla^2 h_{ab} 
- R_{acbd} h^{cd} + R^c_{(a} h_{b)c}
+ \nabla_{(a} \nabla^c h_{b)c}
= - {\textstyle{\frac{1}{2}}} \Delta_L h_{ab}
\ee
often called the Lichnerowicz operator.

\end{enumerate}

The simplest way to perturb the brane system is to take
a GN system, in which the brane stays at $z=0$:
\be
g_{zz} = -1 \;\;\;\; g_{z\mu}=0 \; .
\ee
The remaining gauge freedom allowed is
\be
\xi ^z = f(x^\mu) \;\;,\;\;\;\; 
\xi ^\mu = \int a^{-2} f_{,\nu} \eta^{\mu\nu} + \zeta^\mu(x^\mu)\,.
\ee

We can now input the purely 4D perturbation into the Lichnerowicz 
operator, and after some algebra, the perturbation equations reduce to:
\bea
a^{-2}\left [ a^2 \left ( a^{-2} h^\lambda_\lambda \right )' \right ]'
&=& -\frac{16\pi G_5}{3}\delta(z) a^{-2} {\cal T}^\lambda_\lambda \label{RSptS} \\
\left [ a^{-2} \left ( h^\lambda_{\lambda,\mu} - h^\lambda_{\mu,\lambda}
\right ) \right ] '  &=& 0  \label{RSptV}\\
a^{-2} \partial^2 h_{\mu\nu} - a^{-2} \left [ a^4 \left ( a^{-2} h_{\mu\nu}
\right )' \right ]' - 2 a^{-2}{\bar h}^\lambda_{(\mu,\nu)\lambda}&& \nonumber\\
-aa' \eta_{\mu\nu} \left ( a^{-2} h^\lambda_\lambda \right )'
&=& -16\pi G_5 \delta(z) \left [ {\cal T}_{\mu\nu} - {\textstyle{\frac{1}{3}}}
{\cal T}^\lambda_\lambda \eta_{\mu\nu} \right ] \label{RSptT}
\eea
where brane indices are raised and lowered with $\eta_{\mu\nu}$, and we
allow for a matter perturbation confined to the brane:
\be
T_{\mu\nu} = \frac{6k}{8\pi G_5} + {\cal T}_{\mu\nu}\,.
\label{diffem}
\ee
It is easy to see the
RS gauge is only consistent for vacuum perturbations, and that
the zero modes have the behaviour $\sim a^2$ 
(the graviton \cite{RS,others}) and
$\sim a^2 \int a^{-4}$ (the radion \cite{CGR}).

A complete set of solutions to the free equations is readily found to
be $h_{\mu\nu} \propto e^{ip\cdot x} u_m(z) $ with
\be
u_m(z) = \sqrt{\frac{m}{2k}} \; \frac {J_1({\frac{m}{k}})N_2(m\zeta) 
- N_1({\frac{m}{k}}) J_2(m\zeta) } 
{\sqrt{J_1({\frac{m}{k}})^2 + N_1({\frac{m}{k}})^2}}\,,
\ee
where $\zeta = e^{kz}/k$, from which we can construct the Green's function:
\be
G_R(x,x') = - \int\frac {d^4p}{(2\pi)^4} e^{ip\cdot (x-x')} \left [
\frac{ka^2(z)a^2(z')}{ {\bf p}^2 - (\omega + i\epsilon)^2}
+\int_0^\infty\! dm \frac{u_m(z) u_m(z') }{ m^2 + {\bf p}^2
- (\omega + i\epsilon)^2} \right ]\,.
\ee
\FIGURE[ht]{\epsfig{file=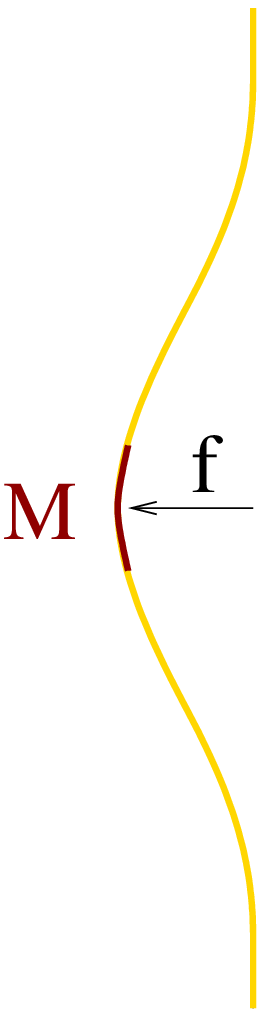,height=8cm}
\caption{
An illustration of the brane bending in response to matter on the brane.}
\label{fig:bbend}}       % Give a unique label
This has the structure of a zero mode (the part proportional to 
$a^2$), and a continuum of KK states. This is seen more clearly
by looking at the restriction to a perturbation in the brane 
induced by a particle on the brane, for which 
\be
G_R(x,0,x',0) = kD_0(x-x') + \int_0^\infty \! dm \; u_m(0)^2 D_m(x-x')\,.
\ee
However, we have to remember that the RS gauge is only 
consistent in the absence of sources;
in the presence of sources we have to fix the trace of the perturbation
to satisfy the Lichnerowicz equation. Strictly speaking, we take
the general metric perturbation, $h_{\mu\nu}$, and decompose
it into its irreducible components with respect to the 4D Lorentz
group (see \cite{CGKP}). This allows for a tensor (TTF) mode, 
a vector, and two scalars in general:
\be
h_{\mu\nu} = h^{\tt TT}_{\mu\nu} + A_{\nu,\mu} + A_{\mu,\nu} +
\phi_{,\mu\nu} - \frac{1}{4} \eta_{\mu\nu} \partial^2 \phi +
\frac{h}{4} \eta_{\mu\nu} \, . \label{decomps} 
\ee
On shell, it can be shown that this reduces (up to purely 4D
gauge transformations) to the following expression
\be
h_{\mu\nu} = h^{\tt TT}_{\mu\nu} - \frac{1}{k} f_{,\mu\nu}
+ 2k a^2 f \eta_{\mu\nu}\,,
\ee
which physically corresponds to the TTF 4D tensor $h^{\tt TT}$, and
a scalar, $f(x^\mu)$, which can be interpreted as a bending of the brane
with respect to an observer at infinity \cite{GT} 
(see figure \ref{fig:bbend}). This brane bending
term couples to the trace of the energy momentum perturbation on the
brane via (\ref{RSptS}), which implies a 4D equation for $f$:
\be
\label{bbendf}
\partial^2 f = \frac{8\pi G_5{\cal T}^\lambda_\lambda}{6} \qquad
\Rightarrow \qquad
f = 8\pi G_5 \int D_0(x-x') \frac{{\cal T}^\lambda_\lambda(x')}{6}\,.
\ee

Solving (\ref{RSptT}), and pulling all this information together, 
we can now write the solution on the brane:
\be
h_{\mu\nu} = -16\pi G_5 \int G_R(x,0;x',0) [{\cal T}_{\mu\nu} 
- {\textstyle{\frac{1}{3}}} {\cal T} \eta_{\mu\nu}]
+ 2k \eta_{\mu\nu} \int D_0(x-x') \frac{8\pi G_5{\cal T}^\lambda_\lambda}{6}
\ee
At mid to long range scales on the brane, the zero-mode dominates
the integral and so we get:
\be
h_{\mu\nu} = -16\pi G_5 k \int D_0(x-x') [{\cal T}_{\mu\nu} 
- {\textstyle{\frac{1}{2}}} {\cal T} \eta_{\mu\nu}]\,.
\ee
Thus, if we identify $G_N=G_5k$ as the 4D Newton constant, we have precisely
4D perturbative Einstein gravity with the correct tensor structure.

The effect of the massive KK modes on the Newtonian potential is also
easily extracted using asymptotics of Bessel functions:
\be
u_m(0) = \sqrt{\frac{m}{2k}} \frac{ J_1\left ( \frac{m}{k} \right )
N_2\left ( \frac{m}{k} \right ) - J_2\left ( \frac{m}{k} \right )
N_1\left ( \frac{m}{k} \right ) }{ \left [ J_1^2\left ( \frac{m}{k} \right )
+ N_1^2\left ( \frac{m}{k} \right ) \right ] ^{1/2} } 
\;\; \sim \;\; - \sqrt{\frac{m}{2k}} \;\;\;\;\; m/k \ll 1 \,.
\ee
To see how these feed into corrections to Einstein gravity, 
consider the effect of a point mass source ${\cal T}_{00} 
\sim M \delta(z) \delta^{(3)}({\bf r})$:
\be
h_{\mu\nu} = -16\pi G_N \int \left ( D_0(x-x') [{\cal T}_{\mu\nu} 
- {\textstyle{\frac{1}{2}}} {\cal T} \eta_{\mu\nu}]
+ \int_0^\infty\! \frac{mdm}{2k^2} D_m (x-x') [{\cal T}_{\mu\nu} 
- {\textstyle{\frac{1}{3}}} {\cal T} \eta_{\mu\nu}] \right )\,,
\label{Newtcorr}
\ee
giving
\be
h_{tt} = - \frac{2G_N M}{r} \left ( 1 + \frac{2}{3k^2r^2} \right )
\qquad , \qquad
h_{ij} = - \frac{2G_N M}{r} \left ( 1 + \frac{1}{3k^2r^2} \right )
\delta_{ij} \,.
\label{RSlingrav}
\ee
Note this is in homogeneous gauge; transforming to the
area gauge (where the area of 2-spheres is $4\pi r^2$) we have
to leading order in $G_NM$:
\be
ds^2 = \left(1-\frac{2G_N M}{\hat r} - \frac{4G_N M}{3k^2{\hat r}^3} \right)
dt^2 - \frac{d{\hat r}^2}{\left (1-\frac{2G_N M}{\hat r} 
- \frac{2G_N M}{k^2{\hat r}^3} \right )} - {\hat r}^2 d\Omega^2\,.
\label{RSlinSch}
\ee
We can visualize RS gravity as lines of force spreading out from the
brane, but being ``pushed back'' by the negative bulk curvature.
At small scales, the lines of force leave the brane and gravity is
5D and weaker. At larger scales, the bulk curvature bends the lines
of force back onto the brane, and so gravity returns to being a 4D force
law. (See figure \ref{fig:rsgrav}.)
\FIGURE[ht]{\epsfig{file=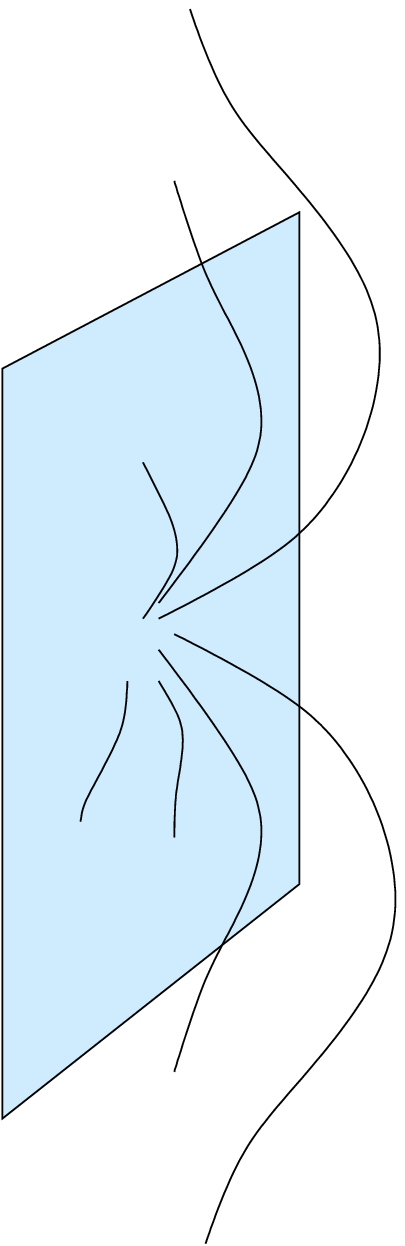,height=8cm}
\caption{A representation of the lines of force for RS gravity.}
\label{fig:rsgrav}}       % Give a unique label

This is weak gravity, but what about strong gravity, such as black holes
or cosmology?

\subsection{Cosmology}

For a cosmological brane, we have to ask whether there are surfaces
of lower dimensionality which have the interpretation of an expanding 
universe. Recall that in standard cosmology, homogeneity and isotropy give
a simple model of the universe in which everything depends on a single
scale factor $a(t)$.
\be
ds^2 = dt^2 - a^2(t) d{\bf x}_\kappa^2
\ee
where the spatial metric $d{\bf x}_\kappa^2$ is a surface of 
constant curvature $\kappa = 0, \pm 1$, and in which $a(t)$ satisfies
a simple first order Friedman equation
\be
\left ( \frac{\dot a}{a} \right )^2 + \frac{\kappa}{a^2} =
\frac{8\pi G_N}{3} \rho
\ee
where $\rho$ is the energy density of the universe, typically modelled
by a perfect fluid with some equation of state $p = w \rho$.

For the cosmological braneworld, homogeneity and isotropy will still
imply a constant curvature spatial universe, but now our ``scale factor''
must depend not only on time, but on the distance into the bulk. The
remaining part of the metric in the $\texttt{t,z}$ directions 
can be made conformally
flat (any two dimensional metric can always be written in a conformally
flat form) and so we may write the overall geometry as \cite{BCG}:
\be
\label{bulkcosmet}
ds^2=e^{2\nu(\texttt{t,z})}(B(\texttt{t,z}))^{-2/3}
(d{\texttt t}^2-d{\texttt z}^2)-B^{2/3}
\left [ \frac{d\chi^2}{1-\kappa\chi^2}+\chi^2 d\Omega_{I\!I}^2 \right]\,.
\ee
The rationale for this specific way of writing the scale factor
becomes apparent once the Einstein equations are analyzed. Here,
$\texttt{z}$ is representing the bulk coordinate away from the brane,
though it no longer corresponds to proper distance. The brane sits
at $\texttt{z}=0$, which can always be chosen to be the 
location of the brane.  (The conformal transformation
${\texttt t}' \pm {\texttt z}' = {\texttt t} \pm {\texttt z} \pm
\zeta({\texttt t} \pm {\texttt z})$ maintains the form of the metric
while taking an arbitrary wall trajectory
${\texttt z}' = \zeta({\texttt t}')$ to ${\texttt z} = 0$.)

In addition,
the presence of a cosmological fluid will alter the usual brane relation
$\E = \T$ by adding additional energy, $\rho$, to $\E$, and subtracting 
pressure, $p$, from the tension $\T$. Thus our brane energy momentum will
now be
\be
T^a_b = \delta (z)\, {\rm diag} \,\left( \E+\rho, \E-p, \E-p, \E-p, 0 \right)
\ee

If we now compute the bulk Einstein equations, the reason for writing
the metric in the slightly unusual form (\ref{bulkcosmet}) becomes
apparent. Using the lightcone coordinates
\be
\label{light}
x_-=\frac{{\texttt t}-{\texttt z}}{2},\qquad 
x_+=\frac{{\texttt t}+{\texttt z}}{2}
\ee
the bulk equations are:
\bea
\label{einstein2}
B_{,+-} &=& \left (2\Lambda B^{1/3}  - 6\kappa B^{-1/3} \right )
e^{2\nu}\label{wave11}\\
\nu_{,+-} &=& \left ( \frac{\Lambda}{3} B^{-2/3} + \kappa B^{-4/3} \right)
e^{2\nu}\label{wave12}\\
B_{,\pm} \left[ln(B_{,\pm})\right]_{,\pm} &=& 2\nu_{,\pm} B_{,\pm} 
\label{inteq}
\eea 
This system is completely integrable, giving, after a change of coordinates,
the bulk solution\footnote{Although note that there is a special case
$B=1$, $2\Lambda=6\kappa$, which is a near horizon limit of a black hole
metric and a critical point of the Einstein equations \cite{ESU,NBH}.}
\be
ds^2 = \left ( \kappa - \frac{\Lambda}{6} r^2 - \frac{\mu}{r^2} \right )
dt^2 - \left ( \kappa - \frac{\Lambda}{6} r^2 - \frac{\mu}{r^2} \right )^{-1}
dr^2 - r^2 d{\bf x}_\kappa^2
\ee
which is clearly a black hole solution. The parameter $\mu$ is 
related to the mass of the bulk black hole via \cite{MP}
\be
M=3\pi\mu/8G_5\,.
\ee
The change of coordinates results in a shift of the brane to 
\be
r = R(\tau)
\ee
where $\tau$ is the proper time of a brane observer.

Thus our cosmological brane is a slice of a black hole spacetime
\cite{BCOS,GCOS,Gubser,BCG}. 
We can think of our brane as moving in the bulk, and as it moves
through a warped background, the brane will experience contraction
or expansion as the surrounding geometry contracts or expands. The
Israel equations give the dynamical equations for the brane trajectory
$R$, which can be massaged into the familiar Friedman form:
\be
\left (\frac{{\dot R}}{R} \right )^2 + \frac{\kappa}{R^2}
= \frac{[8\pi G_5(\E+\rho)]^2 + \Lambda}{36} + \frac{\mu}{R^4}
\ee
(see \cite{BCG} for details).  For a critical RS brane, which has
$\E=6k/8\pi G_5$, $\Lambda = -6k^2$ this gives
\be
\left (\frac{{\dot R}}{R} \right )^2 + \frac{\kappa}{R^2}
= \frac{8\pi G_N\rho}{3} + \frac{(8\pi G_5\rho)^2}{36} + \frac{\mu}{R^4}
\ee
As might have been expected from the calculation of linearized gravity, 
the dominant form of this equation for small $\rho$ is indeed the 
standard Friedman equation. The effect of the brane shows up in 
the $\rho^2$ corrections, dubbed the {\it non-conventional} cosmology 
of the brane \cite{GCOS}. But most interesting from the point of view
of these lectures is the presence of the last term, which is directly
a result of the bulk black hole. This term, proportional to the mass
parameter, takes the same functional form as a radiation source on
the brane. Of course, the presence of the bulk black hole induces a
periodicity in time in the Euclidean section, or, a finite temperature
for any quantum field theory in the spacetime. Computing the background
Hawking temperature of the black hole gives
\be
T_H = \frac{\sqrt{\kappa^2 + 4\mu k^2}}{2\pi r_h}
\ee
where $r_h$ is the location of the event horizon, given by
\be
k^2 r_h^2 = \frac{1}{2} \left [ \sqrt{1+4\mu k^2} -1 \right]\,.
\label{schadsrad}
\ee
For the case
of the RS model, for which $\kappa=0$, this gives
\be
\frac{\mu}{R^4} = \frac{(\pi T_H)^4}{k^6R^4} = \frac{(\pi T)^4}{k^2}
\label{darkrad}
\ee
where $T$ is now the comoving temperature on the brane. That this 
is suggestive of the Stefan law, $\rho \propto T^4$, is not a
coincidence, and is a theme we will
pursue in the next section.

\section{Black holes and holography}
\label{sec:adscft}

Both the linearized gravity result for an isolated mass and the 
brane cosmology metric suggest a somewhat deeper importance to
braneworlds and black holes. The corrections to the Newtonian potential
(\ref{Newtcorr}) in fact coincide precisely with the 1-loop corrections to the
graviton propagator \cite{Duff,DL}, and the cosmological dark radiation
term in the brane Friedman equation corresponds (up to a factor) to the
energy density of a conformal field theory at the Hawking temperature 
of the black hole \cite{Gubser}. 
These clues, and analogies with lower dimensional branes, have led 
to the black hole {\it holographic conjecture} of Emparan,
Fabbri and Kaloper \cite {EFK} which states, loosely speaking, 
that if we have a classical solution to the RS model then we can interpret
the braneworld as a quantum corrected 4D spacetime. In the case of the
black hole, this would mean that we have a quantum corrected black hole.

The reason for putting forward such a conjecture is based in the
adS/CFT conjecture \cite{MAL} of string theory. In string theory,
D-branes arise as the physical manifestation of open string 
Dirichlet boundary conditions. These D-branes are tangible objects 
carrying mass, Ramond-Ramond
charge, and with worldvolume gauge theories to support the 
string endpoints \cite{DBI}. Further, the supergravity solutions
which correspond to the mass and charge of a particular type 
of D-brane must describe the same objects.
The metric for a stack of $N$ coincident D3-branes is given by:
\be
ds^2 = \left(1 + \frac{4\pi gN \alpha^{\prime2}}{r^4} \right)^{-1/2} dx^2_{||}
- \left(1 + \frac{4\pi gN \alpha^{\prime2}}{r^4} \right)^{1/2} dx_\perp^2
\label{D3stack}
\ee
where $g$ is the string coupling and $\alpha'=l_s^2$ the string length scale.
$dx^2_{||}$ and $dx_\perp^2$ are the cartesian metrics of
the spaces respectively parallel and perpendicular to the brane, and
$r$ is the radial coordinate in this latter space.
We trust this supergravity solution in regions where the spacetime curvature
is small, i.e.\ $L^2 \gg \alpha'$, where $L$ is the ambient spacetime
curvature. Obviously, this will be true at large $r$ in (\ref{D3stack}), 
however, at large $r$ the effect of the branes is negligible. 
In order to trust the supergravity solution
in regions where it is nontrivial, i.e.\ 
where $r\sim(4\pi gN\alpha^{\prime2})^{1/4}$, we require $gN \gg 1$.
In this case, we can effectively ignore the ``$1$'' in the prefactor, 
and (\ref{D3stack}) is approximately:
\be
ds^2 = \alpha' \left [ \frac{(r/\alpha')^2}{\sqrt{4\pi gN}} dx^2_{||}
- \frac{\sqrt{4\pi gN}}{r^2} dr^2 - \sqrt{4\pi gN} d\Omega^2_V \right ]\,.
\ee
This metric is adS$_5 \times S^5$.
Thus, if we integrate over the $S^5$, and identify
\be
L = k^{-1} = (4\pi gN)^{1/4} l_s
\label{adslength}
\ee
as the adS length scale, we can directly relate the near horizon 
r\'egime of a stack of D3-branes with the RS model. Further, the
5D Newton constant will be given in terms of the 10D Newton constant
and the volume of the 5-sphere by 
\be
\hbar G_5 =\frac{\hbar G_{10}}{V_5} = \frac{g^2 \alpha^{\prime4} (2\pi)^7}
{16\pi^4 L^5} = \frac{\pi L^3}{2N^2}
\label{hnsquared}
\ee
Thus we can relate finite and {\it classical} quantities in our
5D Einstein theory, such as the adS curvature scale, $L$, and the 
gravitational constant, $G_5$, to quantum mechanical quantities 
such as $\hbar$, and $N$, the number of D-branes. Indeed, we can
potentially take a classical limit, $\hbar\to0$, keeping our
adS scale finite by simply simultaneously taking $N\to \infty$.
On the other hand, this stack of $N$ D3-branes has a low energy $U(N)$ 
worldvolume conformal field theory, and taking $N\to\infty$
corresponds to the t'Hooft limit of the gauge theory. Since
we have set $\hbar \to 0$, on the string side $\alpha'\to0$ ensures
that only this low energy sector remains. This is the
essence of the adS/CFT correspondence, that certain strongly
coupled conformal field theories are dual to string theory on 
certain anti-de Sitter spacetimes.

What does this mean for the RS model? As Gubser first noted,
\cite{Gubser}, brane cosmology with a black hole in the bulk has
the appearance of a radiation cosmology from the brane perspective.
From the Hawking temperature of the bulk black hole, the dark
radiation term has the form (\ref{darkrad}), $(\pi T)^4L^2$.
On the other hand, calculating the energy of a CFT at finite 
temperature (at weak coupling) gives:
\be
\rho = 2 \pi^2 c T^4
\label{weakcoupcft}
\ee
where $c = \hbar(N^2-1)/4$ is the coefficient for the trace anomaly
in super Yang-Mills theory.  Thus as $N\to\infty$,
\be
\frac{8\pi G_N \rho_{CFT}}{3} = \frac{4}{3} \frac{G_NL}{2G_5} (\pi T)^4 L^2
\ee
Clearly, if we identify $G_N = 2G_5/L$, then we see
that the classical bulk black hole has the effect on the brane
of a thermal CFT at the comoving Hawking temperature of the black
hole up to the conventional strong/weak coupling factor of $4/3$. 
Note that the factor of $2$ in the definition of the
4D Newton's constant is due to the fact that in adS/CFT we
have a bulk on only one side of the boundary, whereas in 
physical braneworlds, we have bulk on each side of the brane. This
effectively halves the brane tension, which is the key factor
in the relation between the brane and bulk gravitational
constants. 

This rather physical picture of the interplay between the RS model
and the adS/CFT conjecture is further fleshed out by the 
work of Duff and Liu \cite{DL}, who note that the linearized
corrections to the graviton propagator, calculated in (\ref{RSlingrav}),
precisely agree with the 1-loop linearized corrections to flat space
for a central mass \cite{Duff}. These results are extremely suggestive
that a fully nonlinear classical brane/bulk black hole solution would,
from the brane point of view, correspond to a quantum corrected black
hole. Indeed, it was this perspective that first led Tanaka to 
conjecture that a braneworld black hole must therefore be time-dependent,
to agree with the thermal Hawking radiation from a Schwarzschild 
black hole \cite{TAN}. Emparan, Fabbri and Kaloper  then
pointed out that the issue of time dependence is linked to the 
choice of quantum vacuum, and gave a comprehensive analysis of
3D brane black holes, together with options for the RS
black hole.

Roughly, the picture is as follows. If we consider a closed universe
with a bulk black hole, then the brane is precisely equidistant from
the bulk black hole, and the radiation on the brane is precisely thermal.
However, we could imagine displacing the brane slightly, which would
introduce an inhomogeneity in the dark radiation on the brane. 
Moving one side of the brane even closer to the bulk black hole
would then increase this distortion, and would (hopefully!) correspond
to a collapsing shell of warm radiation on the brane. This could
then form its own black hole, which from the bulk perspective would
correspond to the brane actually touching the black hole. The brane
would remain glued to the black hole for a while, but eventually would
separate, which process would correspond to black hole radiation
(see figure \ref{fig:tdepbh}).
\FIGURE[ht]{\epsfig{file=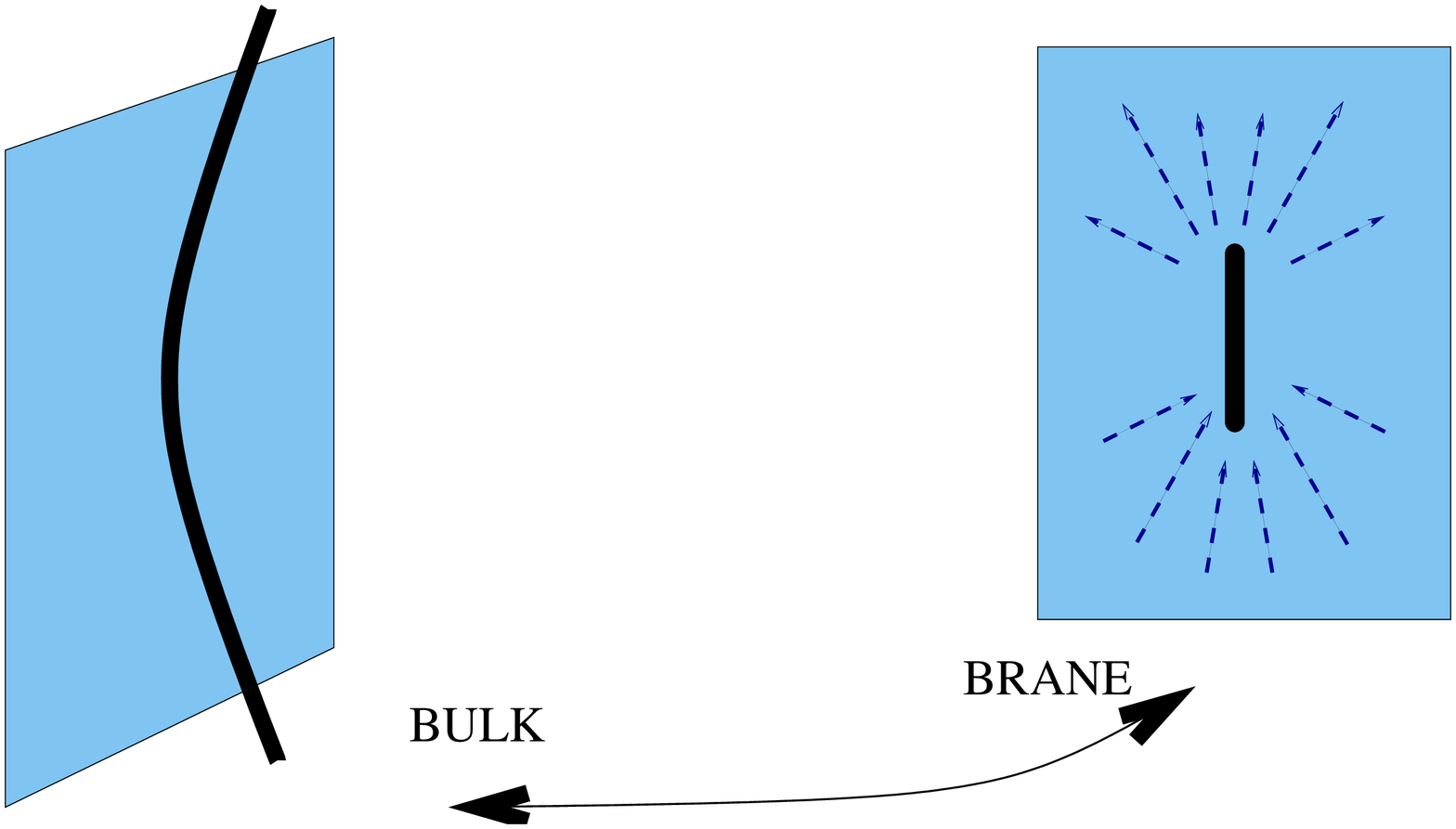,width = 12cm}
\caption{
A cartoon of the time dependent radiating black hole from both the
brane and the bulk perspective. A bulk black hole moves towards the
brane, touches, then eventually recoils back into the bulk. From the
brane perspective this corresponds to anisotropic radiation steadily
accreting around a central point which finally forms a black hole, 
persists for a while radiating, then finally evaporates.}
\label{fig:tdepbh}}       % Give a unique label
%\begin{figure}
%\sidecaption[t]
%\includegraphics[scale=.45]{bend2.eps}
%\caption{An illustration of the curved brane and the Gaussian Normal
%coordinate system. The brane is the solid line at $z=0$, moving out
%a uniform distance from the brane gives a new surface at $z=1$. The
%normal to the brane ${\bf n} = \partial / \partial z$ is indicated.
%As we move from $z=0$ to $z=1$, distances along the brane will change
%in general. This is reflected in the extrinsic curvature (\ref{extcurv}).}
%\label{fig:bentbrane}       % Give a unique label
%\end{figure}

On the other hand, it is always possible that there does exist a
static black hole solution, which asymptotically has the form of
(\ref{RSlinSch}). Such a black hole would, according to EFK,
necessarily have a singular horizon.
This classical solution would be a 5D version of the C-metric
\cite{CMET}, which is a solution representing two black holes
accelerating away from each other. The black holes are being
accelerated by two cosmic strings, one for each hole, which
pull the black hole out to infinity. The exact purely gravitational
solution has a conical deficit which can be smoothed out
by a U(1) vortex \cite{CSD}, rendering
the spacetime nonsingular apart from the central singularities 
of the black holes. It is then straightforward to slice this 
spacetime with a brane \cite{EHM,EGS} thus producing
a 3D braneworld with a black hole. A positive tension brane
retains the bulk without the cosmic string, hence these braneworld
black holes do not need any further regularization. It may
seem strange that a static black hole on the brane is accelerating,
but it is no more unusual than the fact that we are in an accelerating
frame on the surface of the Earth. Geodesics in the 
RS bulk actually curve away from the brane:
\be
2kz(t) \sim  \ln (1+k^2t^2)
\ee
thus any observer glued to the brane is necessarily in an 
accelerating frame.

Moving up one dimension however changes the picture completely. The
mathematics of the pure gravitational equations is now no longer
amenable to analytic study, and no known C-metric exists. Even
higher dimensional ``cosmic strings'' now have codimension three,
and are strongly gravitating \cite{CPB} with potentially singular
geometries.
We will now look at this problem in more detail.

\section{Black hole metric}
\label{sec:bhst}

The first attempt to find a black hole on an RS brane was that of
Chamblin, Hawking and Reall (CHR) \cite{CHR}, in which they replaced
the Minkowski metric in (\ref{rsmet}) by the Schwarzschild metric
(indeed, we can replace $\eta$ in (\ref{rsmet}) by
{\it any} 4D Ricci-flat metric):
\be
ds^2 = a^2(z) \left [ \left ( 1 - \frac{2M}{ r} \right ) dt^2 -
\left ( 1 - \frac{2M}{ r} \right )^{-1} dr^2 - r^2 d\Omega^2_{I\! I} \right]
-dz^2
\label{CHRmet}
\ee
This is the only known exact solution looking like a black hole
from the brane point of view. Unfortunately, it does not correspond
to what we would expect for a brane black hole. If matter is confined
to our brane, we would expect that any gravitational effect is
localized near the brane. For a collapsed star, we would also
intuitively expect that while the horizon might well extend out into
the bulk, it too should be localized near the brane, and the singularity
should not extend out into the bulk. The problem with the CHR
black string,  (\ref{CHRmet}), is that it extends all the way out
to the adS horizon, moreover, at this surface the black hole horizon
actually becomes singular! 

There is however another, more serious, problem with the CHR black
string, and that is that it suffers from a classical instability \cite{BSINS}.
Black string instabilities were first discovered in vacuum, \cite{GL}, for the 
Kaluza-Klein black string:
\be
ds^2 = \left ( 1 - \frac{2M}{ r} \right ) dt^2 -
\left ( 1 - \frac{2M}{ r} \right )^{-1} dr^2 - r^2 d\Omega^2_{I\! I} - dz^2\,.
\label{KKstring}
\ee
This has a cylindrical event horizon, with entropy $4\pi G_N M^2$.
On the other hand, assuming a KK compactification scale of $L_{KK}$,
a 5D black hole of the {\it same} mass as the string (\ref{KKstring})
has an entropy of $8\sqrt{2\pi L_{KK}G_N}\, M^{3/2} / 3\sqrt{3}$. Thus,
for small enough masses relative to the compactification scale
($G_N M \leq 2L_{KK} / 27\pi$) a standard 5D black hole has higher 
entropy than the string, and thus the string should be either
perturbatively or nonperturbatively unstable. 

The existence of the instability can be confirmed by solving the
Lichnerowicz equation 
\be
\nabla^2 h^{ab} + 2 R_{\;c\;d}^{a\;b} h^{cd} = 0\,.
\ee
There is a subtlety involving the initial data surface, which must
be taken to touch the future event horizon (the black hole generically
forms from gravitational collapse), however, there is
an unstable $s$-mode with the form
\be
h^{ab} = e^{i\mu z} e^{\Omega t} \left [ 
\begin{matrix} H^{tt}(h) & h(r)
& 0 & 0 & 0 \cr
h & H^{rr}(h) & 0 & 0 & 0 \cr
0 & 0 & K(h) & 0 & 0 \cr
0 & 0 & 0 & K/\sin^2\theta & 0 \cr
0 & 0 & 0 & 0 & 0 \cr \end{matrix} \right ]
\ee
(Note, this is written in Schwarzschild coordinates for convenience,
but to check $h$ is small, use Kruskals.)
This mode is physical, since any gauge degree of freedom would have 
to be purely 4D, thus satisfying a massless
4D Lichnerowicz equation, whereas this mode satisfies a massive
4D Lichnerowicz equation. 
The effect of the instability is to cause the horizon to ripple.

For the CHR black string, the presence of the bulk cosmological
constant might be supposed to change the technicalities of this 
analysis, however, the crucial feature of the black string instability
is that it is a purely 4D (massive) tensor TTF mode -- i.e.\ it
satisfies the RS gauge! If we work out the perturbation equations 
for the CHR black string background  they have
the particularly simple form:
\be
\left ( (\nabla^{(4)})^2 h_{\mu\nu} + 2 R^{(4)}_{\mu\lambda\nu\rho}
h^{\lambda\rho} \right ) - \left [ a^4\left ( a^{-2} h_{\mu\nu}\right)'
\right]' = 0 \,.
\ee
This means we can
simply take the standard KK instability and substitute the appropriate
massive $z$-dependent eigenfunction: $h_{\mu\nu} = \chi_{\mu\nu} u_m(z)$, so
that $\chi_{\mu\nu}$ satisfies the equation of motion
\be
\left ( \Delta_L^{(4)} + m^2 \right ) \chi_{\mu\nu} = 0
\ee
where $\Delta_L^{(4)} $ is the 4D Schwarzschild Lichnerowicz operator.
In other words, we have the same 4D form for the instability, but a
different $z$-dependence appropriate to the RS background.
Figure \ref{fig:inst} shows the effect of the instability on the
black string horizon, which now ripples with ever-increasing 
frequency towards the adS horizon. 
\FIGURE{
\includegraphics[height=8cm]{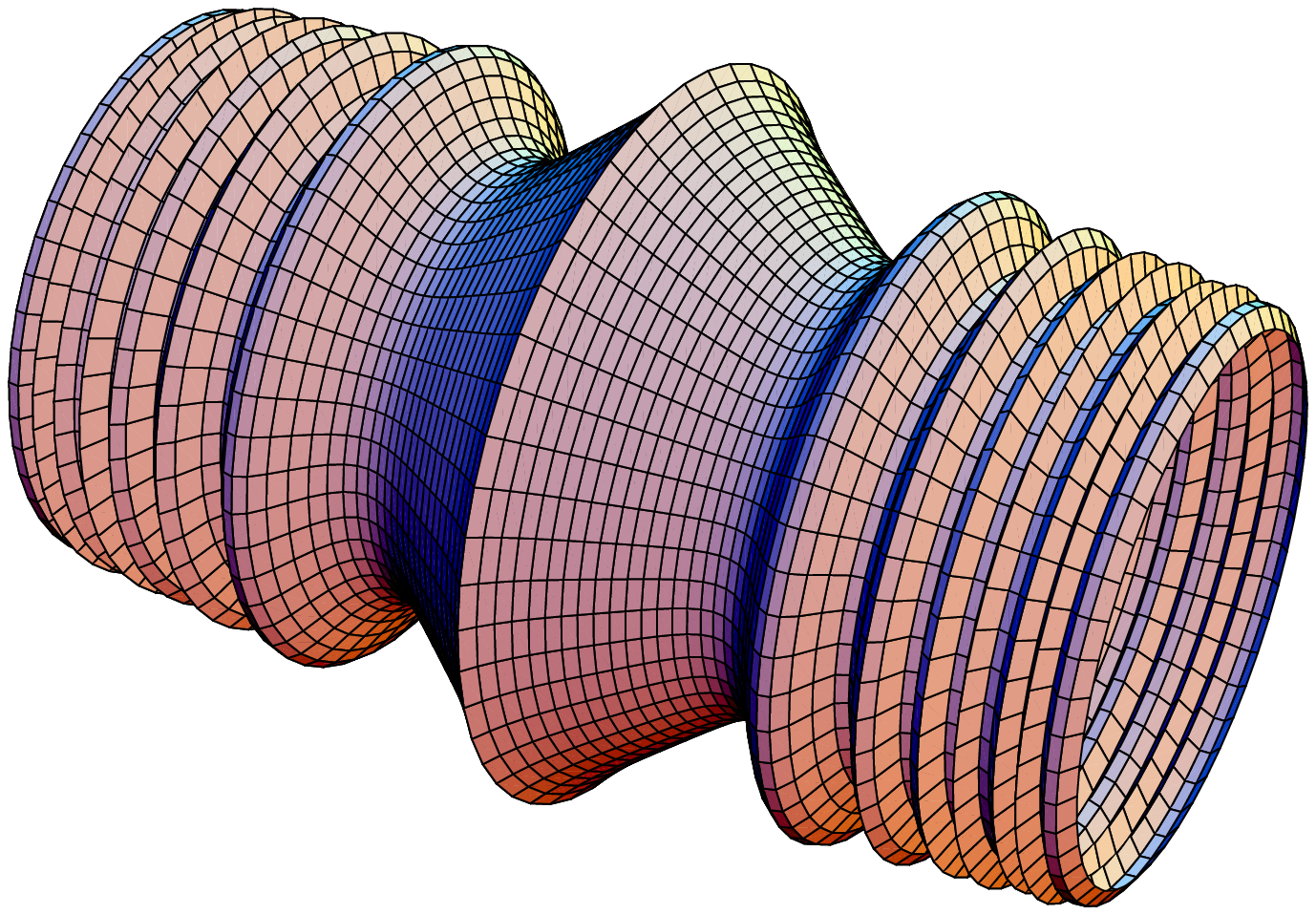}
\caption{The CHR black string horizon after the instability has set in.
The brane is located at the central cusp.}
\label{fig:inst}
}

It is tempting to link the existence of this instability to the
thermodynamic instability of black holes to Hawking evaporation,
however, the timescales have rather different behaviour. Not only
that, but the instability is a dynamical process, and the amplitude
of the instability, ${\cal A}$, is essentially arbitrary. The thermal
radiation from a black hole however, is a quantum process with a 
well defined amplitude. To see the difference, note that for a
black hole emitting radiation into $O(N^2)$ states:
\be
\frac{dM}{dt} \propto \frac{\hbar N^2}{(G_NM)^2} = \frac{L^3}{2G_5(G_NM)^2}
=\frac{1}{G_N} \left ( \frac{L}{G_NM} \right )^2\,.
\ee
For the unstable black string, the mass function on the future
horizon is given by an integral over the KK modes \cite{FabP}
\be
M(v) = M_0 + \int_0^{m_{\rm max}} \frac{dm}{k} \, M_0
(2G_NM_0\Omega - 1/2) \, e^{\Omega v} u_m(0)
\ee
where $v$ is the ingoing Eddington Finkelstein coordinate, and $\Omega$
the half life of the instability at $m$, which is well approximated
by $\Omega(m) = m^2/M - m/2$ (see the plots of $\Omega$ vs.\ $\mu$
in \cite{GL}). Given this approximation, we can compare
the rate of mass loss of the black hole to that by evaporation, by simply
taking $dM/dv|_{v=0}$:
\be
\frac{dM}{dv} = \int_0^{\frac{1}{2G_NM}} 
\frac{dm}{2k} \sqrt{\frac{m}{2k}}(2m^2 - mM - \textstyle{\frac{1}{2}}) \, 
(2m^2 - mM) \propto \frac{1}{G_N} 
\left( \frac{L}{G_NM} \right )^{3/2} + \dots
%\left ( 1 + O(G_NM)^{-2} \right)
\ee
which clearly has a different dependence on $L$ and $M$

It seems therefore that the holographic principle is not so
straightforward to either confirm or implement, and reinforces
the need for an exact solution. A natural method to try would be to take
a similar approach as in cosmology: use the symmetries of the spacetime
and construct the most general metric. Clearly we have spherical symmetry
around the black hole, but we also have a time translation symmetry
(assuming a static solution). This introduces an additional degree of
freedom into the system, which can be parametrized as follows \cite{CG}:
\be
\label{weylmetric}
ds^2 = e^{2\phi/\sqrt{3}} dt^2 - e^{-\phi/\sqrt{3}} \left \{
\alpha^{-1/2} e^{2\chi} (dr^2 + dz^2 ) + \alpha d\Omega_{I\!I}^2 \right \}\,.
\ee
The equations of motion then take the form
\bea
\Delta \alpha &=& -2\Lambda \alpha^{1/2} e^{2\chi-\phi/\sqrt{3}}  
+ 2 \alpha^{1/2} e^{2\chi} \label{alpheq} \\
\Delta \phi + \nabla \phi \cdot \frac{\nabla\alpha}{\alpha}
&=& -\frac{2\Lambda \alpha^{-1/2} e^{2\chi-\phi/\sqrt{3}}}{\sqrt{3}} 
\label{phieq} \\
\Delta \chi + {\textstyle{\frac{1}{4}}} (\nabla \phi)^2 
&=& - \frac{\Lambda \alpha^{-1/2} e^{2\chi-\phi/\sqrt{3}}}{2}
-\frac{\alpha^{-1/2}e^{2\chi}}{2}\label{chieq} \\
\frac{\partial_\pm^2\alpha}{\alpha} + {\textstyle{\frac{1}{2}}} 
(\partial_\pm\phi)^2
- 2 \partial_\pm \chi \frac{\partial_\pm\alpha}{\alpha} &=& 0 \label{consteq}
\eea
where $2\partial_\pm = \partial/\partial(r\pm iz)$.
This is clearly a fairly involved elliptic system, but unlike the 
cosmological equations, it is not integrable. What rendered the
cosmological equations integrable was (\ref{inteq}), of which
(\ref{consteq}) is the counterpart in this set of equations.
The presence of the $(\partial_\pm\phi)^2$ in (\ref{consteq})
means that we can no longer use this to integrate up the other
equations. It is possible to
classify the separable analytic solutions, \cite{CG}, however, none
of these have the form expected of a brane black hole metric. 
The system can of course be integrated numerically, however, 
the typical method appropriate for elliptic systems (relaxation)
is apparently extremely sensitive, and has difficulty dealing with
radically different scales for the black hole mass and the adS
bulk length scales. The consensus seems to be that nonsingular
solutions representing static braneworld black holes exist for
horizon radii of up to a few adS lengths \cite{KU} (see also \cite{BHNUM}).
However, there is no convincing demonstration of the existence
of nonsingular static astrophysical brane black holes.

\section{Approximate methods and solutions for brane black holes}
\label{sec:bhsols}

Since we lack an exact solution, it is natural to attempt approximate
methods to gain understanding of the system. There are two main approaches:
One is to confine analysis to the brane, and to try
to find a self-consistent 4D solution. This has the advantage of
only dealing with one variable (the radius) thereby reducing the problem
to a set of ODE's. However, it has the clear disadvantage that it does
not take into account the bulk spacetime, and therefore will not be 
closed as a system of equations -- inevitably there will be some 
guesswork or approximation involved with terms that encode the bulk
behaviour. The other main approach is to take a known bulk, such as
the Schwarzschild-adS solution, and to explore what possible branes
can exist. Within this method, the branes can either be taken as
{\it probe} branes, i.e.\ branes which do not gravitationally 
back react on the bulk black hole, or as fully gravitating solutions
to the Israel equations, which will therefore have restricted 
trajectories.

Other approaches not reviewed here
include allowing for more general bulk matter,
\cite{bulkbh}, which moves beyond the RS model being considered here.
Also, the extension of brane solutions into the bulk,
have been explored perturbatively \cite{intobulk}, and 
as numerically \cite{initialdata}.

\subsection{Brane approach}

The brane approach is based on the formalism of Shiromizu, Maeda and
Sasaki (SMS) \cite{SMS}, who showed how to project the 5D Einstein-Israel
equations down to a 4D brane system. The SMS method uses the fact
that the RS braneworld has $\mathbb{Z}_2$-symmetry, and writes
(\ref{GCRicci}) at $z=0^+$ using the bulk Einstein equations $R_{ab} = 
4k^2 g_{ab}$ to replace the 5D Ricci tensor, and the Israel equations
(\ref{israeleq}) to replace the extrinsic curvature:
\be
K_{ab}(0^+) = -k \gamma_{ab} + 4\pi G_5 ({\cal T}_{ab} - 
{\textstyle{\frac{1}{3}}} {\cal T} \gamma_{ab})
\ee
using (\ref{diffem}) to define ${\cal T}_{ab}$.
The only term that cannot be substituted by known quantities is 
the contraction of the Riemann tensor. Instead, SMS define an
(unknown) Weyl term:
\be
{\cal E}_{ab} = C_{acbd} n^c n^d = R_{acbd} n^c n^d - \frac{R}{12} \gamma_{ab}
+ \frac{1}{3} ( R_{cd} \gamma^c_a \gamma^d_b - \gamma_{ab} R_{nn})
=R_{acbd} n^c n^d + k^2 \gamma_{ab} \label{SMSweyl}
\ee
which is tracefree.

Using these substitutions, one arrives at a brane ``Einstein'' equation:
\be
~^{(4)}R_{ab} - {\textstyle{\frac{1}{2}}} ~^{(4)}R \gamma_{ab} = 
8\pi G_N {\cal T}_{ab} + \frac{(8\pi G_N)^2}{24k^2} {\cal Q}_{ab} 
+ {\cal E}_{ab}
\label{SMSeqns}
\ee
where the tensor ${\cal Q}_{ab}$ is quadratic in ${\cal T}_{ab}$:
\be
{\cal Q}_{ab} = 6 {\cal T}_{ac}{\cal T}^c_b - 2 {\cal T} {\cal T}_{ab}
- 3 {\cal T}_{cd}^2 \gamma_{ab} + {\cal T}^2 \gamma_{ab}\,.
\ee
Clearly these equations have an attractive simplicity, particularly if
solving for an empty brane, however, it is important to note that the
Weyl term (\ref{SMSweyl}) is a complete unknown, and depends on the
details of the bulk solution.

For the case of the black hole however, one can use a method similar to that
in brane cosmology, \cite{Maa}, 
to decompose the Weyl term into two independent pieces:
\be
{\cal E}_{\mu\nu} = {\cal U} \left ( u_\mu u_\nu 
- {\textstyle{\frac{1}{3}}} h_{\mu\nu} \right ) 
+ \Pi \left (r_\mu r_\nu + {\textstyle{\frac{1}{3}}} h_{\mu\nu} \right )
\ee
where $u^\mu$ is a unit time vector, $r^\mu$ a unit radial vector, and
$h_{\mu\nu} = \gamma_{\mu\nu} - u_\mu u_\nu$ is here the purely
spatial part of the braneworld metric.
This renders the vacuum brane equations (\ref{SMSeqns}) rather similar
to the Einstein equations with a gravitating perfect fluid: the Tolman
Oppenheimer Volkoff (TOV) equations. Of course,
${\cal U}$ and $\Pi$ are complete unknowns, and do not not necessarily
satisfy any conventional energy conditions, however, this notional
similarity is very useful in understanding the physical system, and
in fact allows us to derive useful insight into braneworlds 
stars, \cite{BStars}, such as the fact that the exterior of a collapsing
star is not, in fact, static and Schwarzschild.

The vacuum equations have been solved in many special cases,
for example, Dadhich et.\ al.\ \cite{DMPR}, showed that there was
an exact solution with $\Pi = -2{\cal U}$ having the form of a (zero mass)
Reissner-Nordstrom metric on the brane. Other analytic solutions
can be found by assuming a given form
for the time or radial part of the metric
\cite{CFM,BBH}.
However, a useful approach to solving these equations
is to take an arbitrary spherically symmetric metric,
in which we allow for a general area functional for the 2-spheres,
then apply an equation of state between ${\cal U}$ and $\Pi$ \cite{GWBD}:
\be
\Pi = w {\cal U}\,.
\label{weyleos}
\ee
The Einstein equations reduce to a two dimensional dynamical system
from which it is relatively easy to extract general information about
the system. Obviously, we do not expect that this unknown tensor
will have such a simple equation of state as (\ref{weyleos}), however,
just as in cosmology we approximate the energy momentum of the universe
by various eras with fixed equations of state, it seems reasonable to
approximate the near and far horizon behaviour by a fixed $w$.

At large $r$, we might expect the linearized solution (\ref{RSlinSch})
on the brane, which corresponds to $w=-5/4$. Closer to
the horizon however, it is possible that $w$ could become 
very large and negative. There is in fact an exact solution for 
$w\to -\infty$ which displays features which are
generic to solutions with $w< -2$ (the tidal Reissner Nordstrom
solution):
\be
ds^2 = \left [ (1+\epsilon) \sqrt{1-\frac{2GM}{R}} - \epsilon \right ]^2
dt^2 - \frac{dR^2}{1-\frac{2GM}{R}} - R^2 d\Omega_{I\!I}^2\,.
\label{areaworm}
\ee
This solution has a null singularity at $r=r_1$, the relic of an
horizon, but also note that it actually has a `wormhole', i.e.\
the area of 2-spheres surrounding the origin actually has a minimum
value outside the horizon (for $r_0>r_1$), and is increasing as the
horizon is approached. A sketch of a constant time surface is
shown in figure \ref{fig:worm}.
\FIGURE[ht]{
\epsfig{file=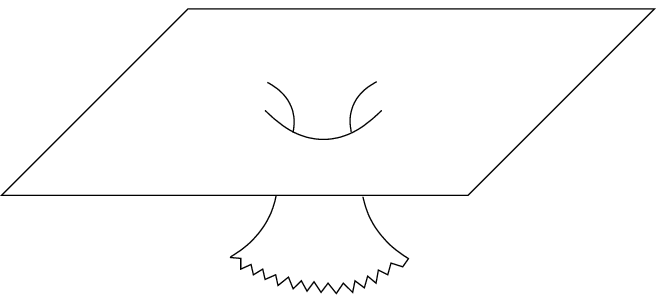,width=9.5cm}
\caption{A sketch of the constant time surfaces of the metric 
(\ref{areaworm}).}
\label{fig:worm}}
It is tempting to conjecture that in fact any such static solution
is singular, but while suggestive, these general arguments are not
a proof, and investigations to determine the nature of the braneworld
black hole horizon have been inconclusive \cite{hornat}.

\subsection{Bulk approach}

The other main approach which can yield insight into brane solutions is
to simplify the problem by taking a known bulk, and exploring the
possibilities for a brane solution with internal spherical symmetry.
With this approach the bulk is now known (although rather rigidly fixed)
and therefore the system has no `unknowns'. The first work in this
area took the brane to be non-gravitating -- a probe brane -- and determined
the general trajectories and dynamics of the brane \cite{PrBr,RECOIL}.
Although this work is not gravitationally self-consistent, it is 
important in particular because it gives insight into highly time
dependent and complicated processes, and is to date the only available
study of the process of a black hole leaving the brane. This has relevance 
for LHC black holes, as the main alternative to decay via Hawking evaporation
is black hole recoil into the bulk, although the holographic point of view
would argue these are indistinguishable \cite{Frolov-recoil}.

In these lectures, we are mostly concerned with the gravitational
properties of brane black holes, and so want to keep the brane at
finite tension and have a consistent back reaction. This is a far
more complicated and restrictive problem, however it is possible to
obtain a linearized metric for a black hole that has left the brane
\cite{grs}. This has the form of a shock wave of spherically symmetric
outgoing radiation on the brane. For a full nonlinear analysis, we
have to look for spherically symmetric branes embedded in a 5D
Schwarzschild-adS spacetime using the Israel formalism. This leads
to some interesting solutions, although the price to be paid is 
that the brane is no longer empty: we require energy momentum on 
the brane to source the gravitational field. This presentation is based 
on \cite{CGKM}, but see also \cite{SS,Galfard}.

The basic idea is to use the Israel equations with a bulk metric
of the general Schwarzschild-adS form. The brane is spherically symmetric,
with additional matter content corresponding
to a homogeneous and isotropic fluid, in other words a braneworld
TOV system. Note however that here there is an actual energy momentum
source on the brane, in addition to the Weyl term (\ref{SMSweyl}) which is
now specified from the brane embedding in the bulk metric:
\begin{eqnarray} \label{eq:metric}
ds^2 &=& U(r)\,d\tau^2 - \frac{1}{U(r)}\,dr^2 -
r^2(d\chi^2 + \sin^2\chi\,d\Omega_{_{I\!I}}^2)\,, \label{general}
\end{eqnarray}
For the brane trajectory, consistent with the $SO(3)$ symmetry, we 
take a general axisymmetric slice $\chi(r)$. Finally, for 
the energy-momentum tensor of the brane we take a general isotropic
fluid source:
\begin{eqnarray}
T_{\mu\nu} = \left[\E (r) - \T(r)\right]\,
u_\mu u_\nu + \T(r)\,h_{\mu\nu}\,.
\label{TOVem}
\end{eqnarray}

It turns out to be convenient to write the Israel equations in terms
of $\alpha =\cos\chi$,
%%%%%%%%%%%%%%
\begin{eqnarray}
Ur \alpha^\prime + \alpha &=&
\frac{8\pi G_5 \E r}{6} \sqrt{[r^2U\alpha^{\prime2}
+ 1-\alpha^2 ]}\label{TOVtt}\\
r^2 U\alpha^{\prime\prime} + \frac{r^2 U'}{2} \alpha^\prime
+ 2 r U \alpha^\prime &+& \frac{r^2 U\alpha^{\prime2}}{1-\alpha^2} 
\left( rU\alpha^\prime + \alpha \right )
\nonumber \\&=& 
\frac{8\pi G_5 \E r}{6(1-\alpha^2)} \left [r^2U\alpha^{\prime2}
+ 1-\alpha^2 \right ] ^{3/2} \label{TOVrr}
\end{eqnarray}
%%%%%%%%%%%%%%
together with a conservation equation which determines $\T$.

These equations can be completely integrated
in terms of a modified radial variable
\be
{\tilde r} = \int \frac{dr}{r\sqrt{U}}
\ee
giving:
\bea
\cos\chi &=& a e^{\tilde r} + b e^{-\tilde r}
\label{gensoln}\\
\E(r) &=& \frac{6}{8\pi G_5 r\sqrt{1-4ab}\,} \left [ \sqrt{U} \left (
ae^{\tilde r} - b e^{-\tilde r} \right ) + a e^{\tilde r} + b e^{-\tilde r}
\right ]\label{genEofr} \\
\T(r) &=& \frac{2}{3} \E(r) + \frac{U'}{8\pi G_5\sqrt{(1-4ab)U}}
\left (ae^{\tilde r} - b e^{-\tilde r} \right) \label{genTofr} 
\eea
Finally the induced metric on the brane is
\be
ds^2 = U d\tau^2 - 
\frac{(1-4ab)dr^2}{U(1-\alpha^2)} - r^2(1-\alpha^2)d\Omega^2
\ee

So far, this is a completely general (implicit) exact solution
which depends on an integral of the bulk Newtonian potential
$U(r)$. Although this is an exact solution, the actual properties of the
brane depend on the specifics of the relation between $\tilde r$ and $r$.
Once this is determined, we have a solution describing a static,
spherically symmetric distribution of an isotropic perfect fluid on
the brane, i.e.\ a solution to the brane TOV system, and therefore
a candidate for a brane ``star''. The extent to which this is a physically
realistic solution will depend on the energy and pressure profiles. 
Note that the profiles $\E(r)$ and $\T(r)$ represent the full
brane energy momentum, and include the background brane tension. The
relevant physical energy and pressure will be defined by
\be
\rho = \E - \E_\infty \qquad, \qquad p = \E_\infty - \T
\ee
where $\E_\infty$ is an appropriate background brane energy, which
has to be identified on a case by case basis.

For a 5-dimensional adS bulk, $U(r) = 1 + k^2r^2$, and the brane satisfies
\begin{eqnarray}
r\cos\chi(r) &=& A\left(\sqrt{U}-1\right) +
B\left(\sqrt{U}+1\right)\,, \label{chiADS}\\
\E &=& \frac{6k^2(A-B)}{8\pi G_5 \sqrt{1-4k^2AB}} \;,\\
\T(r)&=&\E - \frac{2k^2(A+B)^2}{8\pi G_5\sqrt{(1-4k^2AB)U}}\,.
\label{p-AdS}
\end{eqnarray}
where $A= a/k$, and $B=b/k$ in terms of the parameters in 
(\ref{gensoln}).
These brane trajectories are conic sections classified by $|A+B|$. 
For $|A+B|=k^{-1}$, the brane is a paraboloid with critical
RS tension $\E_{RS}$, (\ref{RSTdef}). For $|A+B| > (<) k^{-1}$, 
the brane is an ellipsoid (hyperboloid) with super- (sub-) critical
tension. $A=-B$ is a special case, corresponding to a subcritical
Karch Randall brane and is a straight line.
Figure \ref{fig:ads1} shows sample brane configurations for
various values of the integration parameters $A$ and $B$. 
\FIGURE[ht]{
\epsfig{file=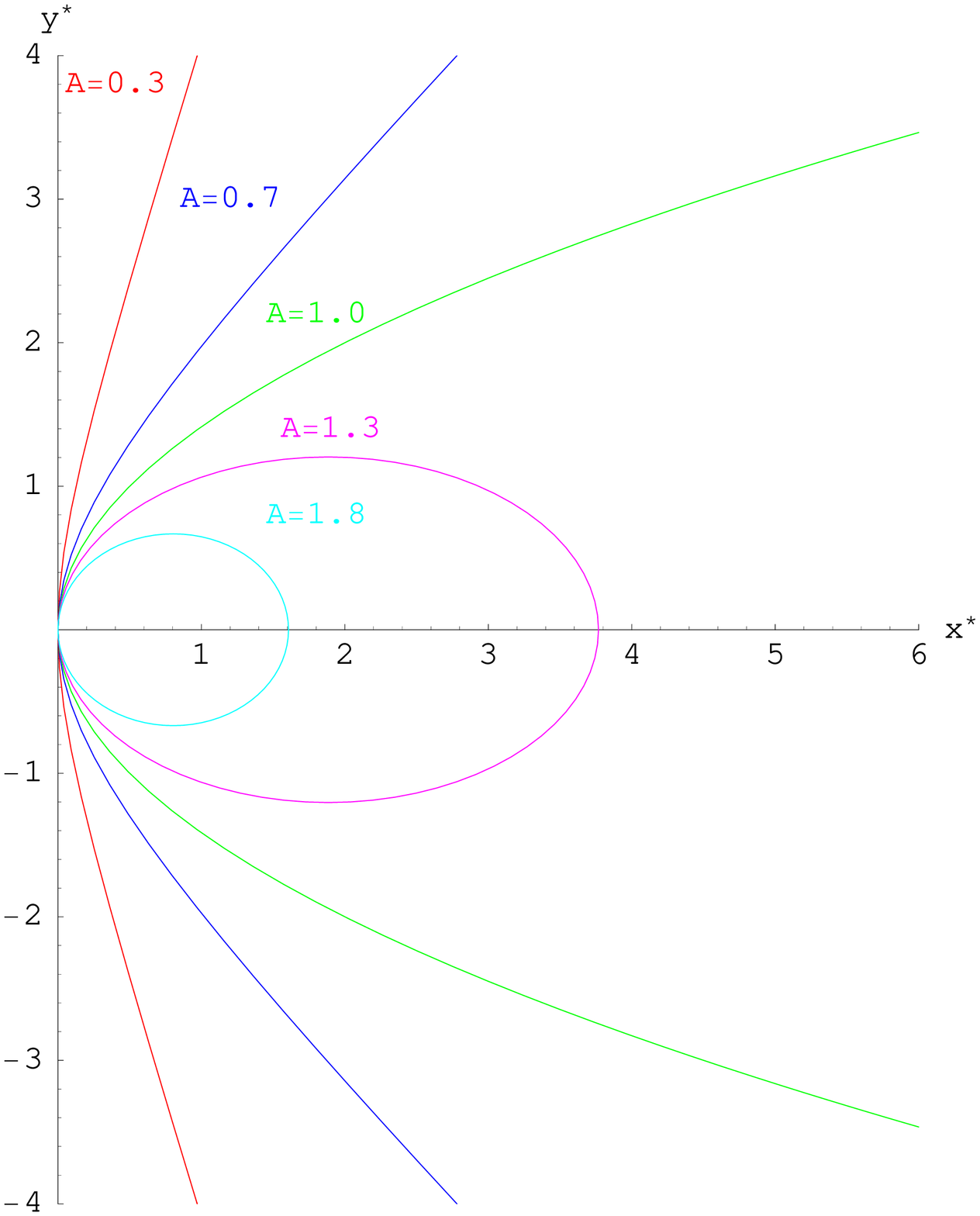,height=9.5cm}
\caption{A selection of branes of varying coefficient $A$, for the
case $B=0, k=1$ in a 5-dimensional anti-de Sitter bulk. }
\label{fig:ads1}}

Notice that the energy density is constant, and requires
$A>B$ to be positive. The tension on the other hand is 
clearly not constant unless $A=-B$.
For the general brane we have a gravitating source composed purely of
pressure!  These branes clearly do not
asymptote exact Randall-Sundrum or Karch-Randall branes.
However, if $|kA|$ and $|kB|$ are large enough, the metric
can be flat (or asymptotically (a)dS) over many orders
of magnitude before the effect of the pressure kicks in.

It is also interesting also to look at a pure Schwarzschild
bulk, $U(r) = 1-\mu/r^2$, for which
\begin{eqnarray}
\cos\chi &=& r\left[A\left(\sqrt{U} - 1\right) +
B\left(\sqrt{U} + 1\right)\right], \label{sol-Schwarz}\\
\E(r) &=& \frac{6}{8\pi G_5 \sqrt{1+4\mu AB}}
\left [ (B-A)(1+U) + 2(B+A)\sqrt{U} \right ] \;,\\ 
\T(r) &=& \frac{2}{8\pi G_5 \sqrt{1+4\mu AB}}
\left [ (B-A)(3+U) + (B+A)(1+3U)/\sqrt{U} \right]
\end{eqnarray}
where now $A = -b/\sqrt{\mu}$, and $B=a/\sqrt{\mu}$ in terms of the
general solution (\ref{gensoln}). Note that by construction, these
trajectories are strictly only valid outside the 
event horizon of the black hole, since the definition of 
the ${\tilde r}$ coordinate involves a branch cut there. 
In contrast
to the adS case, $\E$ is not constant for these branes. For our
solution to correspond to a brane star or black hole,
we require $\E$ to be positive,
and to increase towards the centre of the brane.

Looking at the large $r$ behaviour of (\ref{sol-Schwarz}), we see that
the brane can only reach large $r$ if $B=0$, otherwise the brane is
either a bubble (enclosing the horizon or not, depending on $A$ and $B$)
or an arc touching the horizon. In general, the brane touches
the horizon at a tangent, and the pressure diverges, however, for
one special case $A=-B$, the brane slices through the horizon passing
on to the singularity.
Some sample closed trajectories are shown in figure \ref{fig:schwarz3}.
%%%%%%%%%%%%%%%%%%%%%%%%%%
\FIGURE{
\epsfig {file=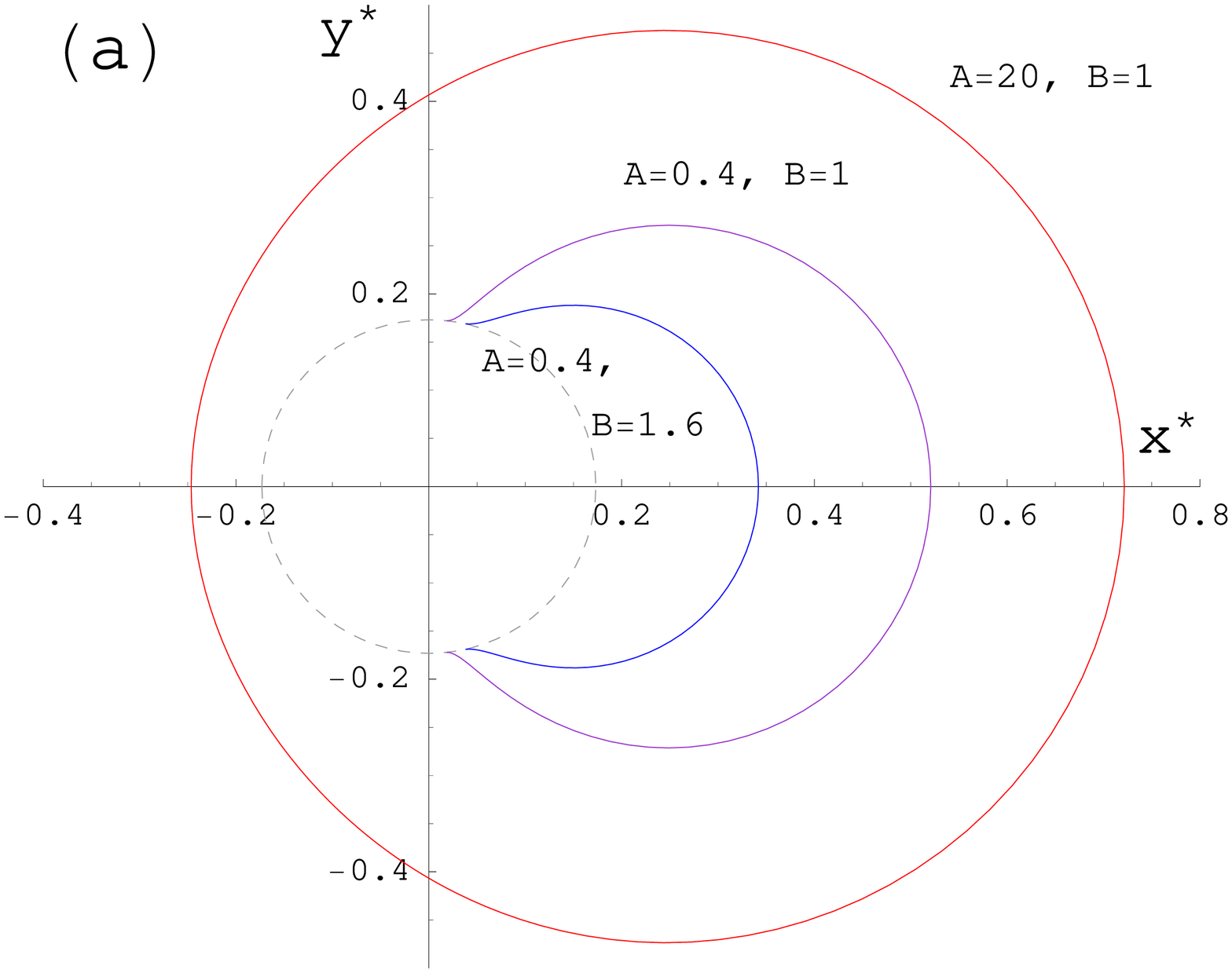, height=6.2cm} \nobreak\hskip1mm\nobreak
\epsfig {file=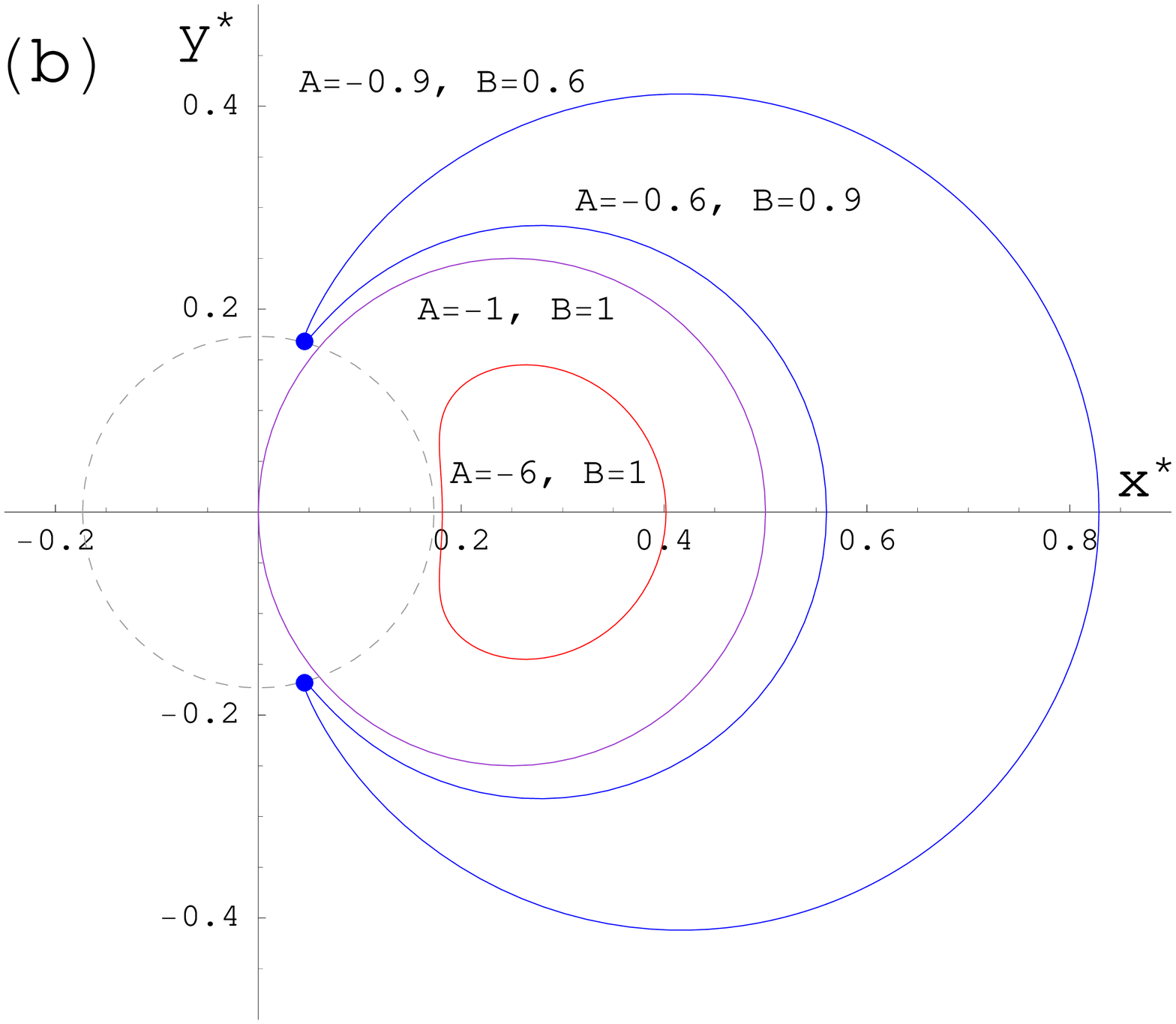, height=6.2cm}
\caption{A selection of branes (solid lines) for the case a) $AB>0$,
and b) $AB<0$ in a 5-dimensional Schwarzschild bulk of fixed mass parameter
$\mu=0.03$. The dashed line denotes the corresponding horizon radius.}
\label{fig:schwarz3}
}
%%%%%%%%%%%%%%%%%%%%%%%%%%%%%%%%%%%%%%%%%%%

The most physically interesting Schwarzschild trajectories are those
which tend to infinity, for which $B=0$, see figure \ref{fig:schwarz1}). 
%%%%%%%%%%%%%%%%%%%%%%%%%%%%%%%%%%%%%%%%%%%
\FIGURE{\hspace*{1cm}
\epsfig{file=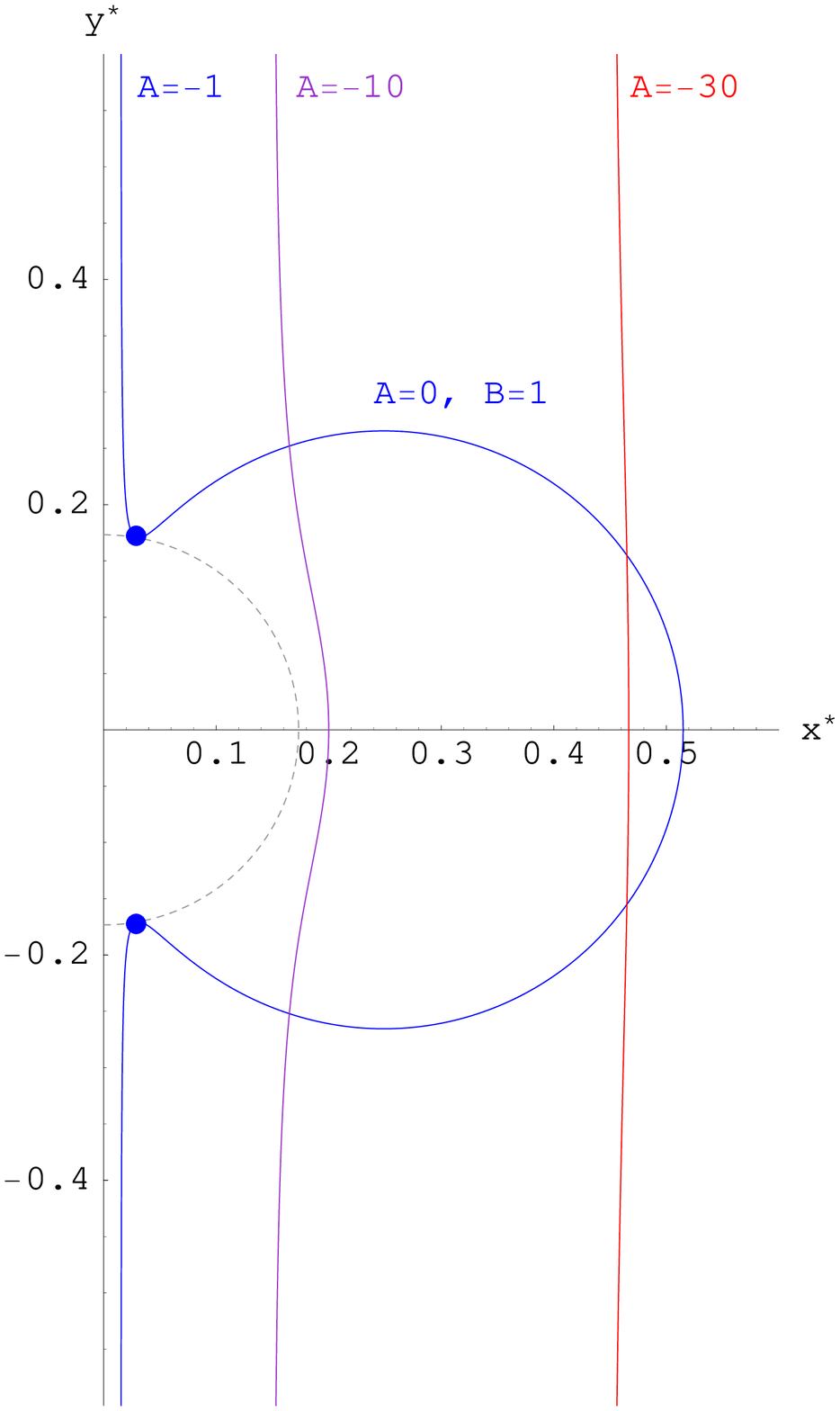,height=10cm}
\caption{A selection of branes for the case $AB=0$, in a 5-dimensional
Schwarzschild bulk of fixed mass parameter $\mu=0.03$. The case $A=0, B=1$
is shown together with a set of branes with $B=0$ and variable $A$.
The dashed line denotes again the event horizon.
}
\label{fig:schwarz1}
}
For these branes $\E_\infty=0$,
and thus
\be \label{schstarem}
\rho = \E(r)= \frac{-6A}{8\pi G_5} \left ( \sqrt{U}-1 \right )^2\,,
\qquad p = -\T = - \frac{\rho}{3} \frac{(\sqrt{U}-1)}{\sqrt{U}}\,.
\ee
For $A<0$, these branes have positive energy and pressure, 
uniformly decreasing as $1/r^4$, and $1/r^6$ respectively.
If $|A|> 1/\sqrt{\mu}$, the brane never touches the horizon and the
pressure remains everywhere finite. Thus these correspond to 
asymptotically empty branes with positive mass sources. Plotting the 
energy and pressure for the brane shows that this does indeed correspond 
to a localized matter source, with the peak energy density dependent 
on the minimal distance from the horizon (see figure \ref{fig:schTOV}).
%%%%%%%%%%%%%%%%%%%%%%%%%%%%%%%%%%%%%%
\FIGURE{
\includegraphics[width=5cm]{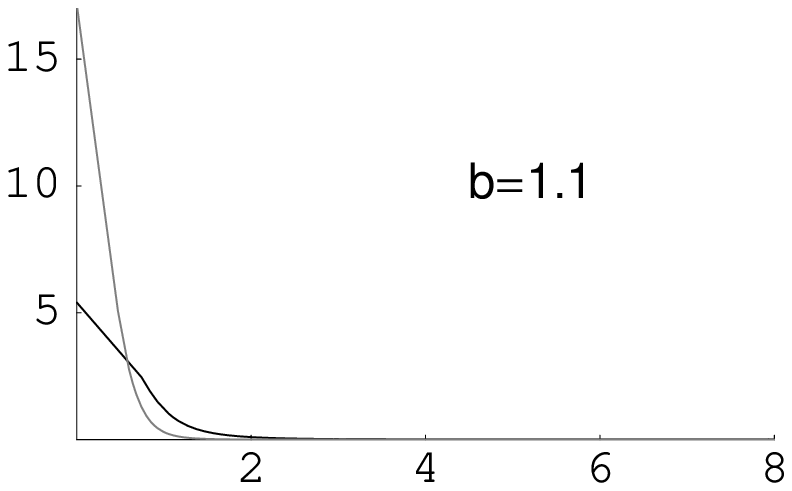}\nobreak\hskip3mm\nobreak
\includegraphics[width=5cm]{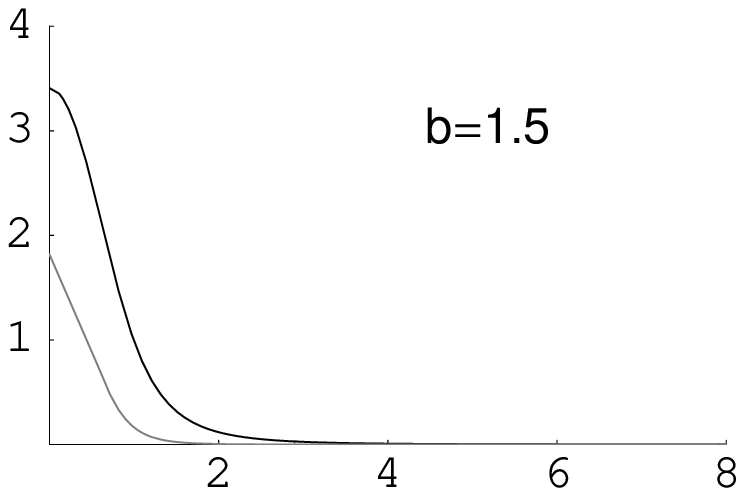}\nobreak\hskip3mm\nobreak
\includegraphics[width=5cm]{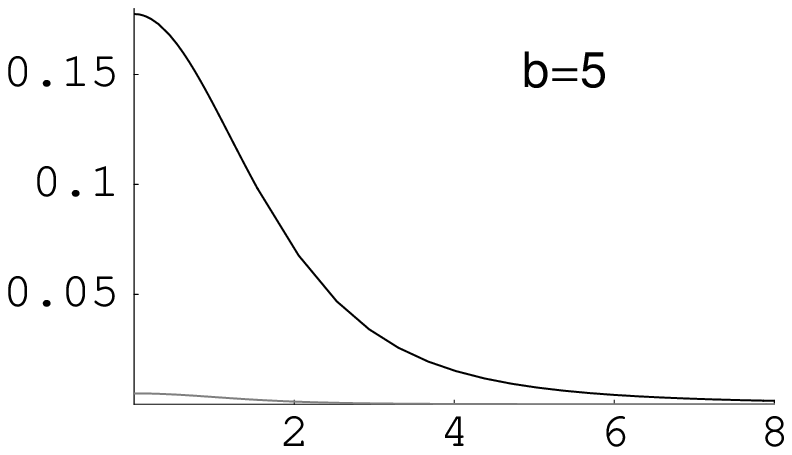}
\caption{The energy (dark line) and pressure (grey line) of
brane stars with a pure Schwarzschild bulk as a function of the
brane radial coordinate ${\hat r}$. The black hole mass
is fixed at $\mu=1$, and the distance of closest approach to
the horizon increases across the plots.}
\label{fig:schTOV}
}
%%%%%%%%%%%%%%%%%%%%%%%%%%%%%%%%%%%%%%%%%%%
The central energy and pressure can be readily calculated from this
minimal radius, $r_m = \mu |A|/2 + 1/2|A|$:
\be
\rho_c = \frac{24|A|}{8\pi G_5(1+\mu A^2)^2}\,, \qquad
p_c = \frac{16|A|}{8\pi G_5(\mu A^2-1)(1+\mu A^2)^2}\,,
\ee
which shows that the central pressure diverges as $\mu A^2\to1$.
This is analogous
to the divergence of central pressure in the four-dimensional TOV
system, which is indicative of the existence of a Chandrasekhar limit
for the mass of the star.

In these spacetimes, there is no actual black hole in the bulk, since
it is the bulk to the right of the brane that is retained. Rather,
it is the combination of the bulk Weyl curvature and the brane bending
which produces the fully coupled gravitational solution.
As the brane moves away from the horizon, the brane matter source spreads
out, but the total mass changes very little, and is determined by
the bulk black hole mass.
The limit on mass is therefore not a true Chandrasekhar limit, but
more a statement about an upper bound on the concentration of matter.
The real reason there is no absolute upper bound is because, unlike the RS
system with an adS bulk, gravity on the braneworld is not localized, nor is
it four-dimensional. Computing the induced metric on the brane
in fact shows that it is the projection of the 5D Schwarzschild metric
on the brane.

\subsubsection{Braneworld Stars : A Schwarzschild-adS Bulk}\label{sec:star}

For the true braneworld star, the appropriate bulk is expected to
be Sch-adS bulk: $U(r) = 1+k^2r^2 - \frac{\mu }{r^2}$. Here
$\tilde r$ has an exact analytic expression
%%%%%%%%%%%%%%
\begin{equation}
{\tilde r} (r) = \frac{1}{k r_h}\, 
{\rm Elliptic\,F}\left[{\rm Arcsin}\left(\frac{r}{r_-}\right),
\frac{r_-^2}{r_h^2}\right]\,,
\label{exact}
\end{equation}
%%%%%%%%%%%%%%
with $r_h$ the black hole horizon, (\ref{schadsrad}), and $r_-$ is defined as
%%%%%%%%%%%%%
\begin{equation}
r_-^2=\frac{-1-\sqrt{1+4k^2 \mu }}{2k^2}\,.
\end{equation}
Since the
Randall Sundrum model is a brane in adS spacetime, we expect that
any consistent brane trajectories in Sch-adS will potentially correspond
to brane stars or black holes. It is worth stressing that these solutions
will not just be brane solutions, but full brane {\it and bulk} solutions,
since the full Israel equations for the brane have been solved in 
a known bulk background. 

From (\ref{genEofr}) the background brane tension is defined as
\be
\E_\infty = \frac{6k(a-b)}{8\pi G_5\sqrt{1-4ab}}\,.
\ee
For large enough $r$, the geometry is 
dominated by the cosmological constant, therefore the
pure adS solutions will be good approximations to any trajectories for
large $r$. Also, if $\mu k^2 \ll 1$, i.e.\ if
the black hole is much smaller than the adS scale,
we expect that in the vicinity of the horizon the Schwarzschild 
solutions will be good approximations for the brane, therefore for
small mass black holes, we might expect brane trajectories to be
well approximated by some combination of Schwarzschild and adS
branes. 
Because the $\tilde r$-coordinate has been zeroed at infinity
(for easy comparison with the pure adS limit)
the range of $\tilde r$ in Sch-adS is finite, and decreases
sharply with increasing $\mu$.
This suggests that trajectories in large mass Sch-adS black hole
spacetimes are more finely tuned, and possibly more restricted than
in small mass black hole spacetimes.

Like adS spacetime, the Sch-adS trajectories can be classified according
to whether they asymptote the adS boundary at nonzero $\chi$, at
$\chi=0$, or do not reach the boundary at all, i.e.\ are closed 
bubbles. These correspond to subcritical, critical,
or supercritical branes ($a+b<1$, $a+b=1$, and $a+b>1$) respectively.
A sample of brane trajectories is shown in figure \ref{fig:schads}.
%%%%%%%%%%%%%%%%%%%%%%%%%%%%%%%%%%%%%%
\FIGURE{\hspace*{1cm}
\includegraphics[width=8cm]{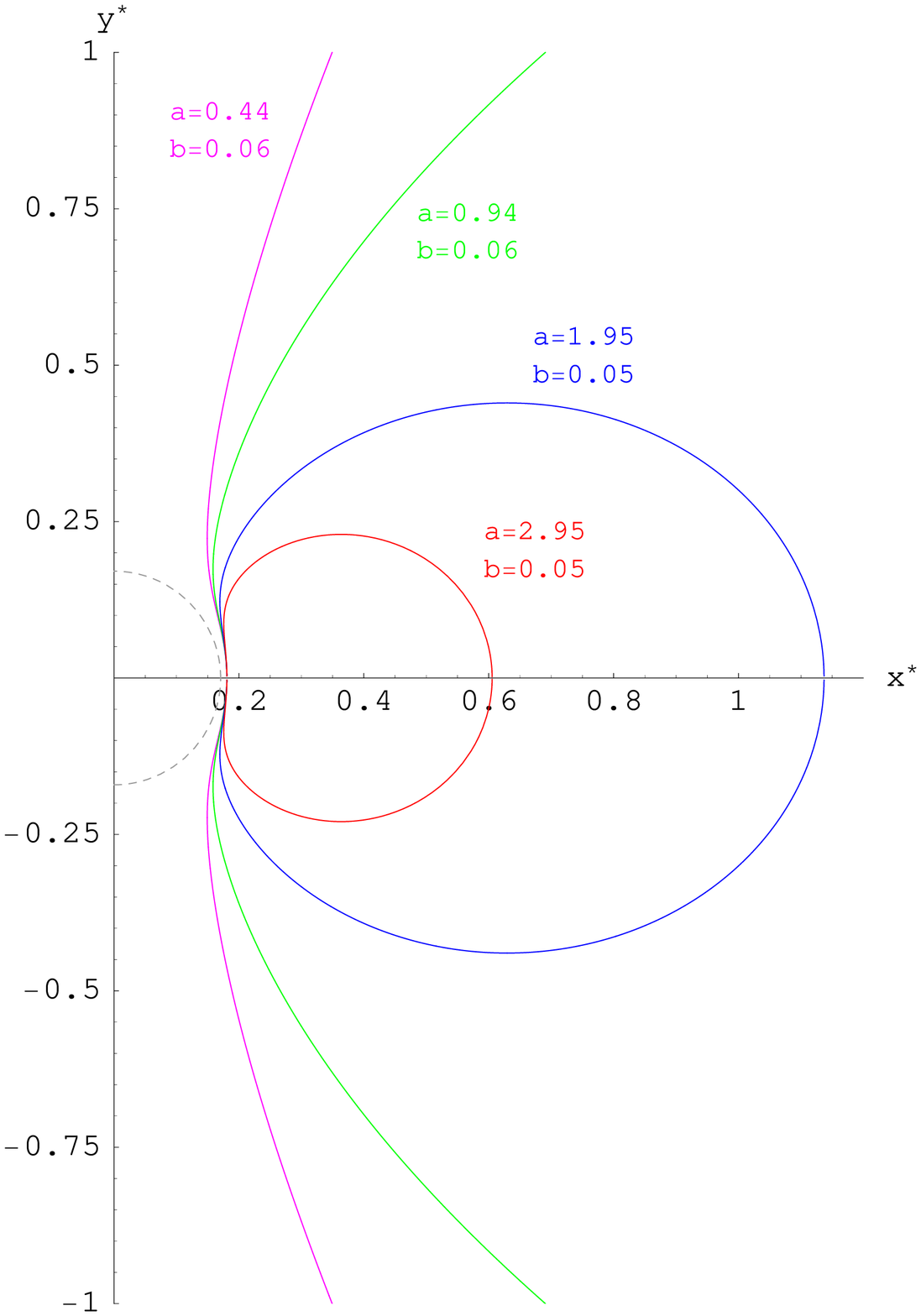}\hspace*{1cm}
\caption{A mixture of brane trajectories in a 5-dimensional
Schwarzschild--anti de Sitter background of fixed parameters $k=1$
and $\mu=0.03$. Note how these are deformed from those of figure
\ref{fig:ads1} by the black hole horizon.}
\label{fig:schads}
}
%%%%%%%%%%%%%%%%%%%%%%%%%%%%%%%%%%%%%%

The supercritical branes are qualitatively similar to the pure
Schwarzschild case, however, it is interesting to note that in
each case there exists a purely empty spherical brane, equidistant 
from the horizon. This corresponds to the Einstein static universe
\cite{ESU}, which from the brane perspective is a closed universe
stabilized by a combination of the cosmological constant (the brane
is supercritical) and the CFT dark radiation term. Using the holographic
intuition, we might expect that by displacing this universe slightly
we could mock up the start of gravitational collapse, however, a quick
computation shows that displacing the brane relative to the black hole
slightly sets up an energy {\it deficit} on the part of the brane closer to 
the black hole!

%%%%%%%%%%%%%%%%%%%%%%%%%%%%%%%%%%%%%%
\FIGURE{
\includegraphics[height=8cm]{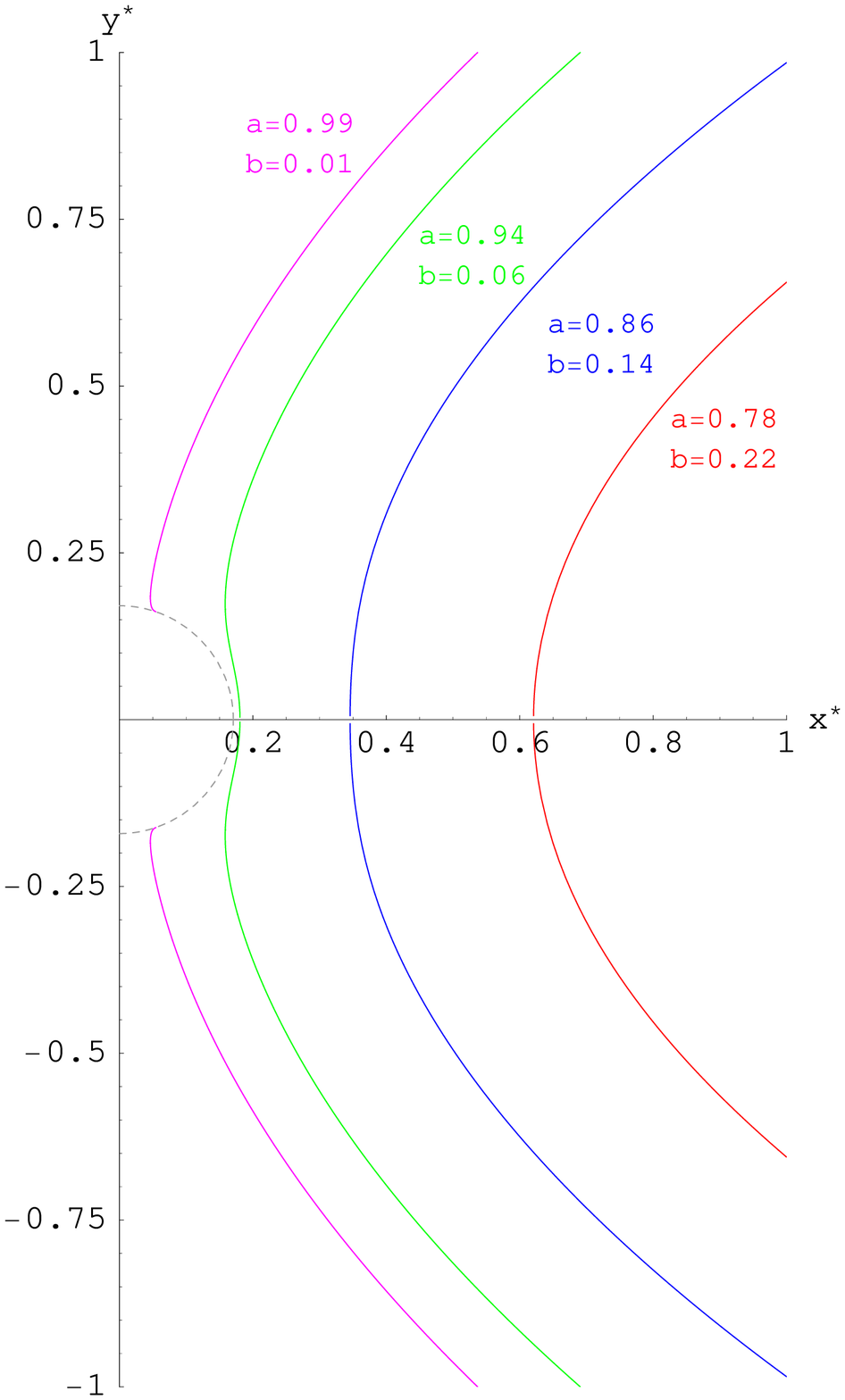}\nobreak\hskip3mm\nobreak
\includegraphics[height=85mm]{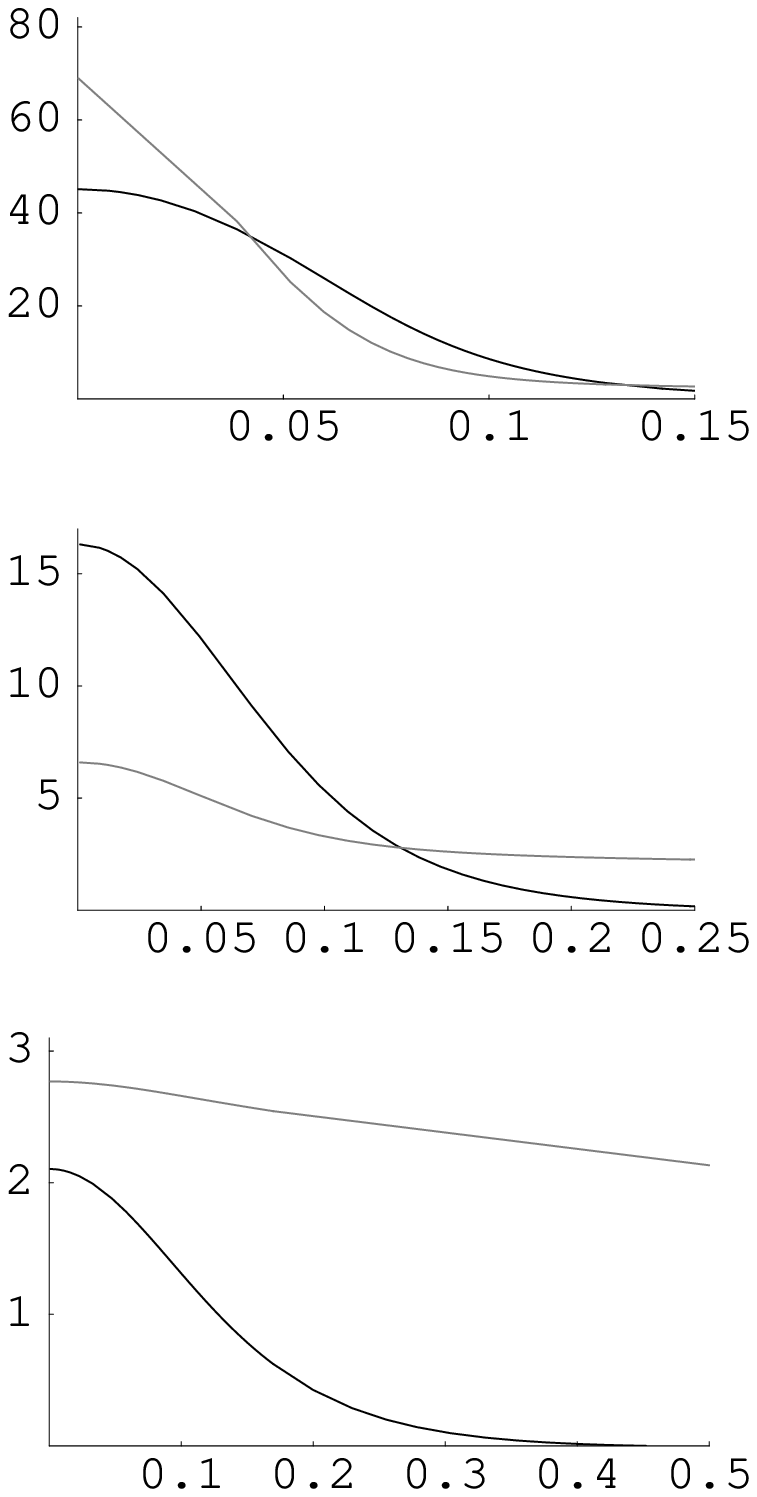}
\caption{(a) A sample of critical brane trajectories with $a+b=1$ in a 
5-dimensional Schwarzschild--anti de Sitter background of fixed
parameters $k=1$ and $\mu=0.03$. The dashed line denotes again the
horizon. (b) A set of plots of the brane energy (black line) and
pressure (grey line) for a sequence of critical branes moving 
away from the horizon.}
\label{fig:crit}
}
%%%%%%%%%%%%%%%%%%%%%%%%%%%%%%%%%%%%%%

In the case of critical branes, $a+b=1$, which means that the 
brane trajectories asymptote the adS boundary at exactly $\chi=0$. 
The branes are thus open, and may or may not touch the black hole 
horizon depending on the exact values of the parameters $a$ and $b$. If
\be
a-b < |\tanh{\tilde r}_+/2|\,,
\ee
the trajectory remains away from the horizon, otherwise it will touch
the horizon and have a pressure singularity.  A sample of
critical trajectories in a Sch-adS background is shown in figure
\ref{fig:crit}(a). 

For branes that 
avoid the horizon the energy density is positive, peaking
at the center, and dropping rapidly to the background value, 
undershooting it very slightly to form an underdense region 
at very large $r$.
The pressure also reaches its maximum value at the center, but
is uniformly decreasing with $r$, at a much slower rate, consistent with
the pressure excess observed for the pure adS branes. Apart from this
pressure excess, the other main difference with pure Schwarzschild 
trajectories, is that the brane matter can no longer universally satisfy the
Dominant Energy Condition (DEC) ($\rho\geq|p|$).
In pure Schwarzschild, the DEC is satisfied except for 
branes which skirt extremely close to the horizon, where the local 
Weyl curvature causes the pressure to diverge. 
This phenomenon is also observed for the Sch-adS branes skimming
close to the horizon, however, as we increase $b$
the central energy dominates the pressure for only a finite range of
$b$ before once again dropping below the pressure. This is because the 
further we move away from the horizon the adS curvature becomes more
important, and for pure adS branes, the effect of the adS curvature is 
to induce a pressure excess.
In figure \ref{fig:crit}(b), the energy density and pressure of
the matter on the brane is shown for a sequence of critical branes in a 
Sch-adS background displaced an increasing distance from the horizon. 

Subcritical branes are largely similar to critical branes, and
correspond to open trajectories that asymptote the adS boundary,
although at nonzero $\chi$ in this case. 
The same bound as before, i.e.\ whether
$|\cos \chi| \simeq |a e^{\tilde r_h} + b e^{-\tilde r_h}| \leq 1$,
will determine whether the brane terminates on the event horizon
or remain on the RHS of it.  The energy density
and pressure profile in this case is again similar to the one found
for critical branes.
Once again, for a large family of parameters $a$ and $b$, solutions
with a positive energy excess at the center of the brane may be
easily found.

One special subcritical trajectory found in the pure adS case was the 
Karch-Randall trajectory, $a+b=0$. We can extend this to Sch-adS 
obtaining
\be
\cos\chi = 2a \sinh {\tilde r}
\ee
however, since $a>0$ for a positive energy trajectory, this has
$(\cos\chi)'>0$, and hence the energy density is always increasing
with $r$. Thus, whether or not these trajectories terminate on the horizon,
they always correspond to energy deficits on the brane, and hence
negative mass sources from the point of view of a brane observer.

To sum up: we can get static solutions to the brane-TOV
equations, and hence static brane stars. Unfortunately, the restricted
form of the bulk leads to unphysical asymptotic behaviour away from the
star in the form of a pressure excess. One possible way of removing this would
be to perturb the bulk slightly at large $r$ to remove this excess. 
However, another interesting route to explore is to make the trajectory
time dependent. In \cite{BStars} it was argued that the spacetime surrounding
a collapsing brane star would be time dependent even though it was vacuum. 
In fact, the RS trajectory {\it is} time dependent when written in global
adS coordinates, which of course are the coordinates used for the 
Schwarzschild-adS metric:
\be
\sqrt{1+k^2r^2} \cos kt - kr\cos\chi = e^{-kz} = 1\,.
\label{RSwallgenu}
\ee
The RS wall is oscillatory because the spherical coordinates are the
universal covering space of adS, and so the `wall' is actually an
infinite family of walls, each in the local patch covered by the
horospherical coordinates. 
Since $r=0$ is a geodesic of the spherical
adS spacetime, the image of $r=0$ in the Randall-Sundrum spacetime,
which is a hyperbola, will be a geodesic in the RS spacetime.
Therefore, if we put a black hole at $r=0$, it should look
like a particle in the RS spacetime, at least to a first approximation.

We can generalize the brane trajectory to $\chi(r,t)$, and compute
the corresponding time dependent versions of (\ref{TOVtt},\ref{TOVrr}),
then find the energy momentum source required on the brane.
The idea is that a time dependent brane solution would describe
a black hole forming from the collapse of radiation, and its 
subsequent evaporation, thus
it is not clear whether we should expect a pure
brane energy momentum solution; rather,
a solution corresponding to the collapse of matter on the brane is
perhaps more physically realistic.
The energy momentum of a surface slicing the Sch-adS
spacetime is given by the Israel junction conditions as:
\be
{\cal T}_{\mu \nu}= \frac{2}{8\pi G_5}\left ( 
K_{\mu \nu} - K h_{\mu \nu}\right ) +\frac{6k}{8\pi G_5}\,h_{\mu \nu} \,.
\ee
Clearly, since the trajectory is time dependent, the energy momentum
will also be time dependent, however, since the largest effect of the bulk
black hole will be represented by the $t=0$ slice of the braneworld
-- the point of closest proximity -- we evaluate the energy momentum at $t=0$. 
For a pure RS trajectory, the black hole
causes the energy of the brane to decrease from its critical
value, whereas both the radial and azimuthal tension increase, thus
the brane matter violates all the energy conditions!  However,
this was not unexpected as the RS trajectory was not modified,
and the main feature of the static brane solutions was that they
responded to the bulk black hole by bending. Indeed, in a
definitive brane gravity paper, \cite{GT}, Garriga and Tanaka
showed that a crucial part of obtaining four dimensional
{\it Einstein} gravity (i.e.\ with the correct tensor structure)
was what could be interpreted as a brane bending term. As shown in
section \ref{sec:RSgrav}, the effect of matter on the brane is 
to ``shift'' the brane with respect to the acceleration horizon in 
the bulk (\ref{bbendf}).
Clearly then, if a black hole forms on the brane, we would expect
the brane to respond to this matter by bending. 

A shift in the position of the brane corresponds to $kz \to 1 +
k\delta z$, and trying a test function:
\be
\cos \chi(t,r) \simeq \frac{1}{r}
\left (\sqrt{1+ r^2}\ \cos\tau - \frac{1}{1-\frac{q}{r^p}} \right )\,,
\ee
gives the behaviour shown in figure \ref{fig:bend} for a range of
of $p$ and $q$. (The brane bending of $1/|{\bf x}|$
corresponds approximately to $p=1/2$.)
%%%%%%%%%%%%%%%%%%%%%%%%%%%%%%%%%%%%%%
\FIGURE{
\includegraphics[width=5cm]{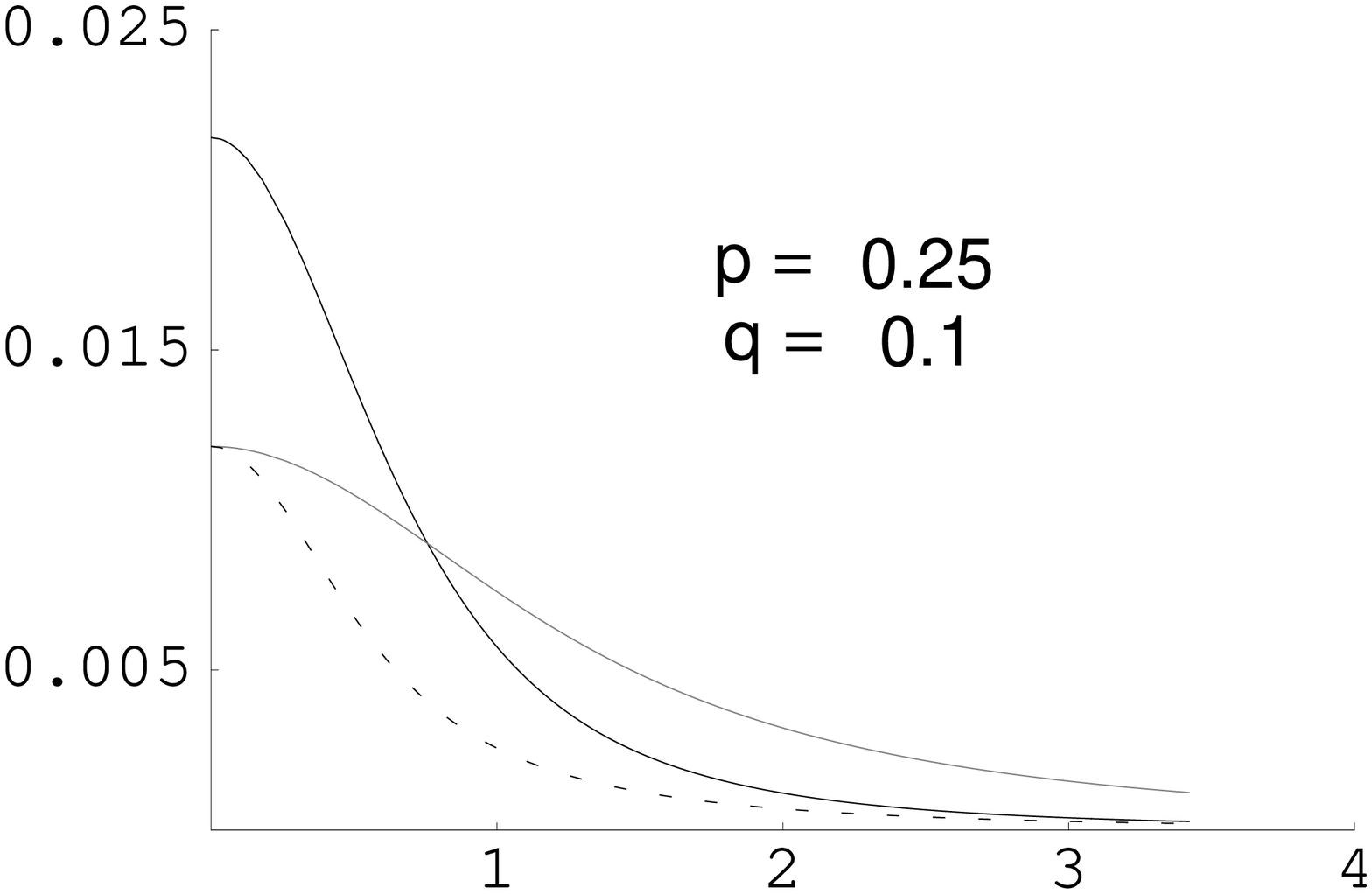}\nobreak\hskip3mm\nobreak
\includegraphics[width=5cm]{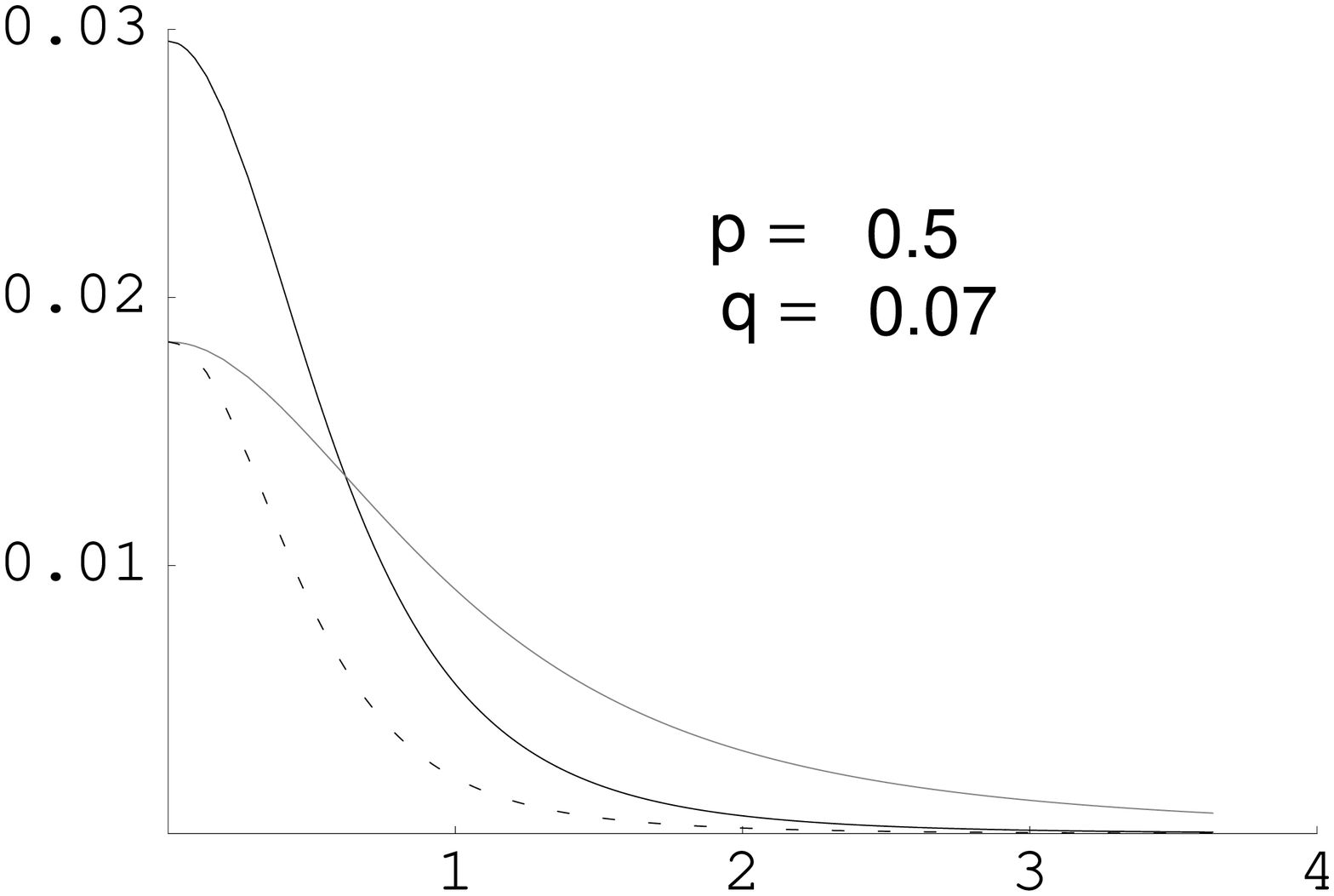}\nobreak\hskip3mm\nobreak
\includegraphics[width=5cm]{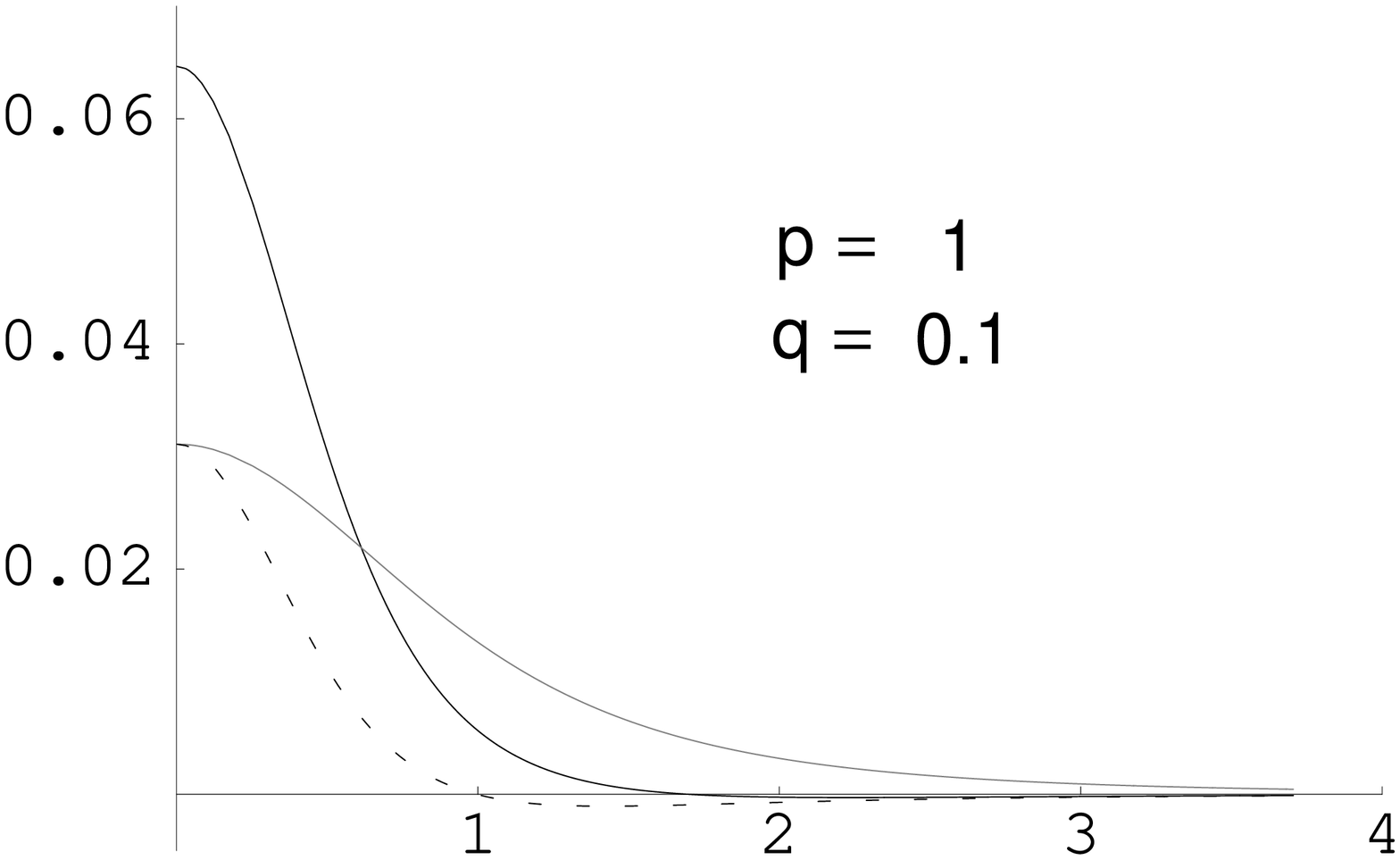}
\caption{A selection of plots of brane energy-momentum with brane bending
included for a range of amplitudes and powers of $r$. The brane energy is
shown in units where $\E_{RS}=3$, and the radial brane distance in units of
$L$. The solid black line is the energy, the dashed line the radial pressure,
and the gray line the angular pressures.}
\label{fig:bend}
}
%%%%%%%%%%%%%%%%%%%%%%%%%%%%%%%%%%%%%%

The brane energy momentum in figure \ref{fig:bend} satisfies the WEC,
however, not the DEC. If the brane is bent instead towards the black hole 
the brane WEC is violated. The excess of angular pressure is somewhat 
similar to the pressure excesses in the static brane trajectory, however,
unlike the static trajectories, here the black hole actually is in the 
bulk, hence these are true candidates for black hole recoil into the bulk.

\subsubsection{The interaction of black holes and branes}

The main motivating factors for obtaining a time-dependent 
braneworld black hole are to gain insight into the 
back-reaction of Hawking radiation on
a quantum corrected four-dimensional black hole, and to 
understand the process of black hole recoil from a braneworld. 
Presumably the time-dependent process will be some perturbed version of a
time-dependent brane trajectory in five-dimensional
Sch-adS spacetime. By allowing the brane to intersect the 
bulk black hole horizon, this would appear to describe 
black hole formation and  evaporation via transport of a 
bulk black hole to the brane, and subsequent departure back into the
bulk. 
When the brane hits the black hole, we might expect
some part of it will be captured
by the black hole, and will therefore remain behind the event  horizon even
when the black hole has left the brane, effectively having been chopped
off from the rest of the brane. This feature is seen in the probe brane
calculations of \cite{RECOIL}, and we expect this to hold in the case of
a fully gravitating brane. In support of this, we can appeal to 
the case of a cosmic string interacting with a black hole, where 
early work indicated that strings would be captured \cite{LM}, and via
self-intersection would leave some part behind in the black hole.
Gravitational calculations of the fully coupled string/black hole
system show explicitly how this ties in with the thermodynamic 
process of string capture and black hole entropy \cite{ABEGK}. 
\FIGURE[ht]{\epsfig{file=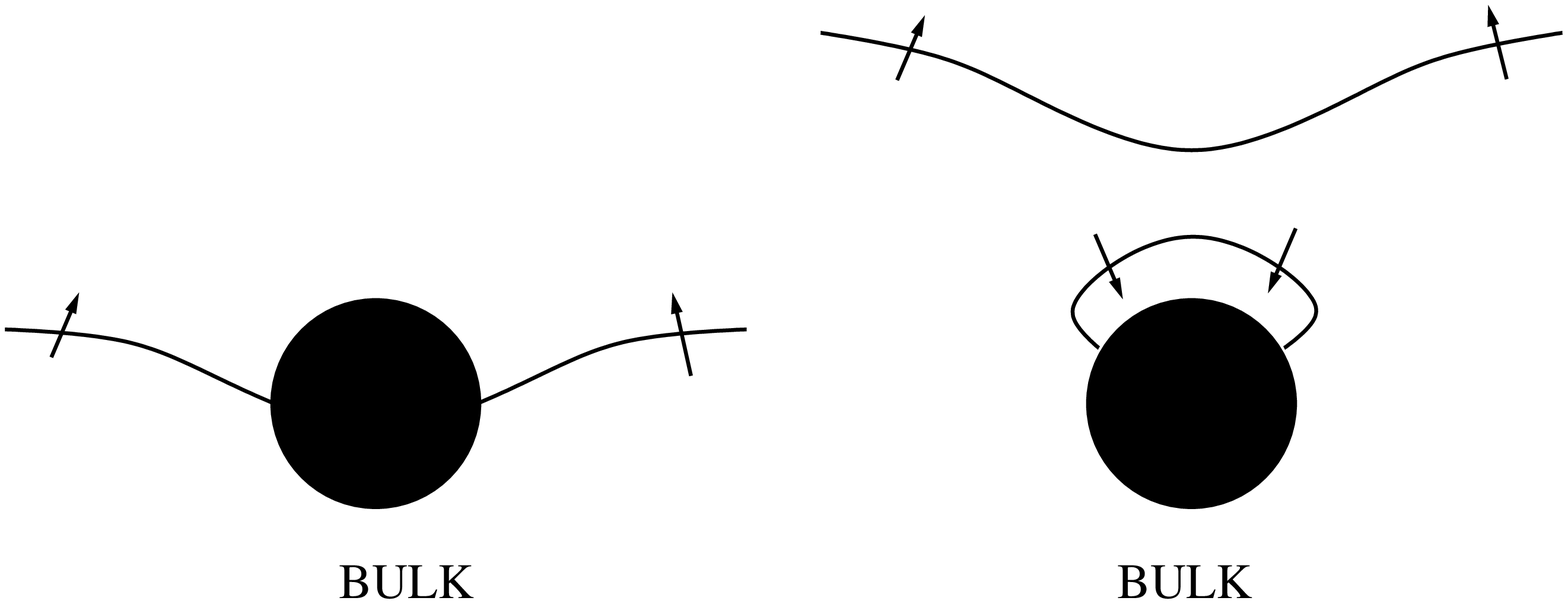,width=12cm}
\caption{
An illustration of brane capture by a black hole. On the left, the
black hole is on the brane, with the brane moving upwards. On the right,
the brane has left the black hole by self intersecting and cutting
off a bubble which falls into the black hole.}
\label{fig:bcapture}}       % Give a unique label

The basic idea is that once part of a brane has fallen into the event
horizon of a black hole, it can no longer leave. Thus, if the brane has
enough kinetic energy to subsequently pull away from the black hole,
the price it must pay is to leave behind the part that has already been 
captured, see figure \ref{fig:bcapture} \cite{RECOIL}. However
, RS braneworlds are {\it not}
probe branes, but are strongly gravitating objects, and therefore any
dynamic process must also be gravitationally consistent. 
From the gravitational point of view, when the black hole captures part
part of the brane and excises it from the whole, the black hole must 
increase in mass. This interplay is seen particularly clearly in the
related case of the cosmic string \cite{ABEGK}, where a cosmic string
piercing a black hole alters the thermodynamic relations between mass,
entropy and temperature. In that case, the (static) results are entirely
consistent with the black hole having captured a length $4G_NM$ of
cosmic string, thus increasing its mass. Just as in the cosmic 
string case, the capture of the codimension 1 RS brane by
the black hole will turn out to be important in establishing the 
thermodynamic viability of the black hole recoil process.

At a first pass, it seems that in fact black hole recoil cannot occur
in RS braneworlds due to a simple entropy argument \cite{nogo}.
In 5D, entropy is proportional to $M^{3/2}$, hence two black holes 
of mass $M/2$ have less entropy than a single black hole of mass $M$. 
However, this argument is both incorrect in the evaluation of 
the entropy, and misses additional contributory factors such as brane 
bending and brane capture by the black hole. 

To get a better estimate, first note that 
entropy is proportional to horizon area/volume, which for
Sch-adS is not simply related to the mass, but also to the adS scale:
\be
{\cal S} \propto 2\pi^2 r_h^3 = \frac{\pi^2}{\sqrt{2}k^3} 
\left ( \sqrt{1+\frac{32G_5Mk^2}{3\pi}} -1 \right )^{3/2}\,.
\ee 
Note that if $G_NM \geq 3.35L$, then the entropy of two black
holes of mass $M/2$ will in fact be {\it greater} than that of
a single black hole of mass $M$.
Therefore, at least from this rather approximate entropic argument, 
black hole recoil would seem to be problematic only for small black
holes. 
On the other hand, in any dynamic process, we must take into account the 
capture of part of the brane by the black hole. Consider the idealized
situation where we have a black hole intersecting the brane along
its equator, in this case, a volume of $4\pi r_h^3/3$ of brane has
been captured by the black hole, with a corresponding mass of
\be
\delta M = \frac{6k}{8\pi G_5} \frac{4}{3} \pi r_h^3
= \frac{1}{2\sqrt{2} G_N} \left [ \sqrt{1+4\mu k^2} \, -1 \right]^{3/2}\,.
\ee
Adding this mass to the recoiled black holes results in an order of 
magnitude improvement to the bound on $M$ coming from the entropy:
for $G_NM \geq 0.35L$, the entropy of the recoiled black  holes
becomes greater than that of the black hole on the brane.

Finally however, the most crucial factor is the brane bending. For
a mass on the brane, the brane bends away from the acceleration horizon,
and (as we have seen) the brane tends to bend away from the black hole.
This effect will be most marked for the smallest black holes. We therefore
have to correct the entropy argument to allow for the fact that more
than half of the black hole horizon is sticking out into the bulk.
(See figure \ref{fig:bcapture}.) Ignoring the effect of the captured
brane increasing the mass, a quick calculation shows that the effective
mass of the intermediate black hole stuck on the brane is
\be
M_{\rm int} = \frac{2\pi M}{2\chi_0 - \sin 2\chi_0}
\ee
where $\chi_0>\pi/2$ is the minimal angle at which the brane touches the
event horizon (assuming the black hole approaches from $\chi=\pi$).
For $\chi_0 > 17\pi/30$, a rather modest amount of brane bending,
the entropy of the recoiled black holes is always greater.

It is important to note that these arguments use the
standard entropy of the isolated Sch-adS black hole. In
other words, they assume a static solution with an event horizon
at $r_h$. Clearly in the time-dependent spacetime there is some
question about whether this approximation is valid, 
and entropy arguments should be used with caution,
nonetheless, for small black holes, where we might expect them to
be more reliable, taking into account brane bending
and fragmentation shows that it is by no means entropically preferred
for a black hole to stick to the brane.

%%%%%%%%%%%%%%%%%%%%%%%%%%%%%%%%%%%%%%%

\section{Outlook}
\label{sec:outlk}

As we have seen, the problem of braneworld black hole solutions
is rather complex, and extremely interesting. The holographic
principle puts forward the tantalizing prospect that if we can
find a classical brane black hole solution (be it time dependent
or static) then this gives us invaluable information about the 
quantum corrected black hole. The failure to find a classical solution
so far can therefore be reinterpreted as the difficulty of 
consistently quantizing gravity. Yet the picture is not quite so clear.
There have been several attempts to solve the brane black hole system
numerically, \cite{BHNUM}, but as yet no unequivocal result. 
As we have seen, finding classical solutions directly is extremely
difficult, and the only progress that has been made is partial,
either by ignoring the bulk, or by relaxing the restrictions on the
brane. 

One interesting possibility, discussed in \cite{FRW}, is that 
the holographic
principle is in fact not applicable to the RS model, and that the lack
of an exact solution is unrelated to any problem of quantum gravity.
Fitzpatrick, Randall and Wiseman (FRW) suggest that it is not 
appropriate to use the
adS/CFT conjecture, as this refers to a quantum field theory at strong
coupling, and the relation between the classical bulk solution and the
quantum corrected brane solution requires the relation (\ref{hnsquared})
where the classical effect is related to the full $N^2$ degrees
of freedom of the field theory. Since the field theory is strongly
coupled, it is not obvious that we will indeed have access to all
the $N^2$ states in all cases. For example, we do not see quarks or
gluons outside the nucleus, so why should we expect to access the full
range of states far away from a black hole? 

Without an exact solution,
there is no way of exploring which of these insights, the holographic
picture of EFK discussed in section \ref{sec:adscft}, or the gluon 
analogy of FRW, is correct. FRW are of the opinion that there does 
exist a nonsingular, static braneworld black hole solution, and proposed
the CHR black string as a counterexample to the holographic conjecture.
The main problem with this solution is that to render it stable a
second brane is required in the bulk. This corresponds to an infra-red
cut-off in the CFT, and it is by no means clear how this additional
complication affects the holographic argument. 

There is however another option for exploring the physics of braneworld
black holes, and that is to move to the Karch Randall set-up \cite{KR}.
The KR brane is slightly de-tuned from the critical RS value, and is 
sub-critical, with an effective negative cosmological constant 
residing on the brane. KR branes are thus adS slicings of adS. 
From the holographic point of view, this complicates the picture, as
we are no longer in the near horizon limit of a stack of D3-branes, 
however, the KR brane can possibly be related to a defect CFT
dual to the intersection of a probe D5 brane with a stack of D3
branes \cite{DCFT}. The advantage of considering this slightly 
detuned situation is that black holes in adS can be thermodynamically
stable \cite{HP}, and therefore the back reaction due to Hawking
radiation can, in principle, be computed. On the other hand, the
adS black string in adS becomes stable once the mass is sufficiently
high \cite{adsstab}, which has been argued to be dual to the 
Hawking-Page transition \cite{CK}. Thus, for large mass black holes
on the KR brane, we can perform a direct
comparison between the strong coupling holographic back reaction,
and the weak coupling Hawking radiation back reaction.

Such a comparison was made in \cite{GRZ} using Page's heat kernel
method \cite{PHK} for approximating the radiation back reaction.
The physical set-up is that we have two KR branes stretching through
the bulk, each with positive tension, and each cutting off the boundary of
adS, hence each providing a UV cutoff CFT. The black string stretches
between the two branes, and for large enough mass is stable (see figure
\ref{fig:KRBS}):
\begin{equation}
ds^2 = \frac{L^2}{\cos^2\theta}
\left [ \left( 1 + k_4^2 r^2 - \frac{2G_NM}{r} \right ) dt^2 
- \frac{dr^2}{\left( 1 + k_4^2 r^2 - \frac{2G_NM}{r} \right )} 
- r^2 d\Omega_{I\!I}^2 - d\theta^2 \right ] 
\end{equation}
where $k_4 = k\cos\theta_0$, with $\pm\theta_0$ being the location
of the KR branes, is the 4D adS curvature scale.
\FIGURE[ht]{\epsfig{file=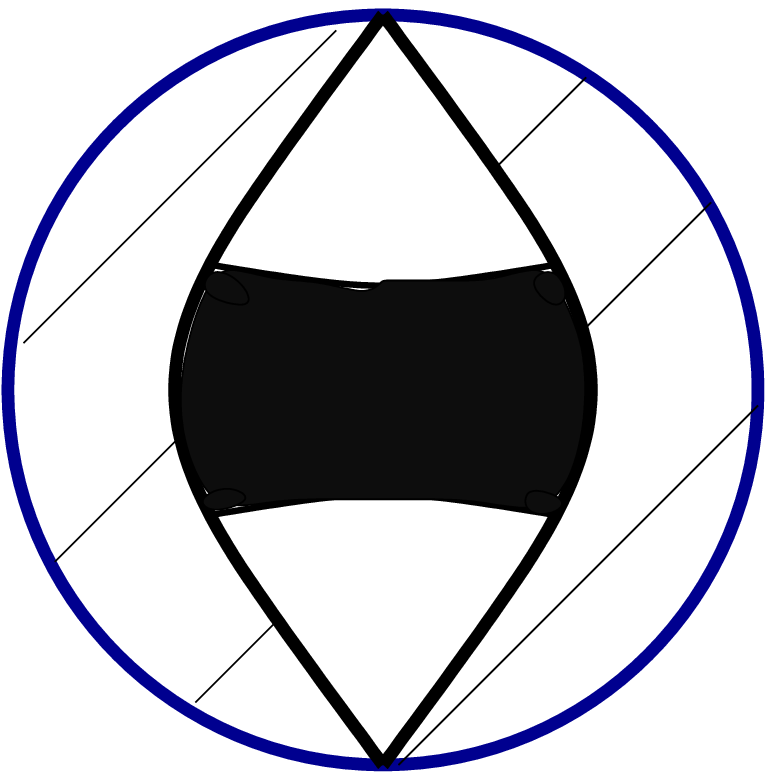,width = 7cm}
\caption{A sketch of the KR black string. The blue circle is the adS
boundary, which is excised from the braneworld spacetime. The string 
goes through the adS bulk between the two KR branes. Because the 
string has finite proper length relative to its mass, it can be stable 
for sufficiently large mass.}
\label{fig:KRBS}}       % Give a unique label

Restricting ourselves to a single brane, the geometry is that of 4D
Sch-adS, and we can perform a standard weak coupling computation of
the energy momentum tensor of the Hawking radiation.
Figure \ref{fig:weakcoup} shows the energy and pressure of the 
thermal bath produced by the black hole (see \cite{GRZ} for details).
Notice how at large $r$ the energy and pressures asymptote the form
of a cosmological constant.

On the other hand, we have a full brane$+$bulk classical solution, and
we can directly compute the effective stress tensor on the brane. 
It is clear before starting however, that this will not have the
form of figure \ref{fig:weakcoup}, as these varying energies and
stresses will backreact on the spacetime to give a modification 
of the Sch-adS solution, whereas the classical solution is pure
Sch-adS. On the other hand, although this is the classical brane
solution, that does not mean that there is no back reaction on
the brane energy momentum. In fact, the correction to the brane
energy momentum is interpreted via the conventional 4D Einstein
equation. From the brane point of view, we are unaware of the extra
dimension, and therefore we interpret any deviation from the 
standard Einstein equation as additional energy momentum. Thus,
while our KR brane energy momentum must have the form of a cosmological
constant, it is possible that this is renormalized from the 
expected bare value.

To see how this works, let the tension of the KR brane be
\be
\E = \frac{6k}{8\pi G_5} + \lambda = \frac{6k\sin\theta_0}{8\pi G_5}
\label{KRtens}
\ee
where $\lambda <0$ is the bare tension on the brane.
On the other hand, the {\it actual} 4D cosmological constant is
given by
\be
\Lambda_4 = -3 k_4^2 = 8\pi G_4 \lambda_{\rm eff} \,.
\label{lambdaphys}
\ee
Note that in this case, the 4D gravitational constant is not labelled 
as $G_N$, since the relation between the brane and bulk gravitational
constant is dependent on the brane tension, not the background adS
curvature \cite{GP}, and is altered from the critical RS relation:
\be
G_4 = \frac{4\pi G_5}{3} \E G_5 \,.
\label{KRG4}
\ee
From the definition of $k_4$ and (\ref{KRtens},\ref{KRG4}), the
value of the bare tension is:
\be
\lambda = \frac{3}{4\pi G_4} \left ( k^2 - k_4^2-k\sqrt{k^2-k^2_4} \right)
\ee
Therefore, since the `expected' value of the cosmological constant
is $8\pi G_4\lambda$, we can compute the correction to the brane
energy momentum as:
\be
\langle T^\mu_\nu \rangle =
\frac{8\pi G_4 \lambda - 3k_4^2}{8\pi G_4} \delta_nu^\mu
= \frac{3(2k^2-k_4^2-2k\sqrt{k^2-k_4^2})}{8\pi G_5\sqrt{k^2-k_4^2}}
\ee
This is the precise (classical) braneworld result. 
We can obtain the holographic renormalization result
\cite{holren}, by taking the limit as the brane approaches the 
boundary, or by approaching the critical RS limit $\lambda \to 0$.
As $k_4\to 0$, we get 
\be
\langle T^\mu_\nu \rangle
= \frac{3k_4^4}{32\pi G_5}
\ee
which agrees with the strong coupling holographic result \cite{GRZ},
up to the expected factor of two which arises from the
braneworld set-up having two copies of the bulk, one on each side of 
the brane.

It is intriguing that the black hole apparently does not radiate 
in the strong coupling picture. 
This is a direct consequence of the fact that the bulk spacetime is foliated
by conformal copies of the Schwarzschild-adS black hole. This
`translation invariance' means that the classical KK graviton modes
are not excited in the background solution, and geometrically the only
possibility is renormalization of the cosmological constant.
It is possible that the black string solution is not the correct
black hole metric candidate, however, one might expect that for
brane black holes with $r_h > L$, there is a unique stable
regular black hole geometry, which this solution is.

Thus, the KR black string provides a counterexample to the 
expectation that a classical braneworld black hole corresponds to
a quantum corrected 4D black hole. There are of course many caveats
to this claim. Clearly the KR brane is not the near horizon limit of
a stack of pure D3-branes, and therefore we do not expect the CFT
to be a simple SYM. However, the fact that the renormalization
of the stress tensor is proportional to $N^2$, yet vanishes in
the critical RS limit is supportive of the arguments of \cite{FRW}.
Obviously this debate is far from over! (See \cite{recent} for
some recent work.)
\FIGURE[ht]{\epsfig{file=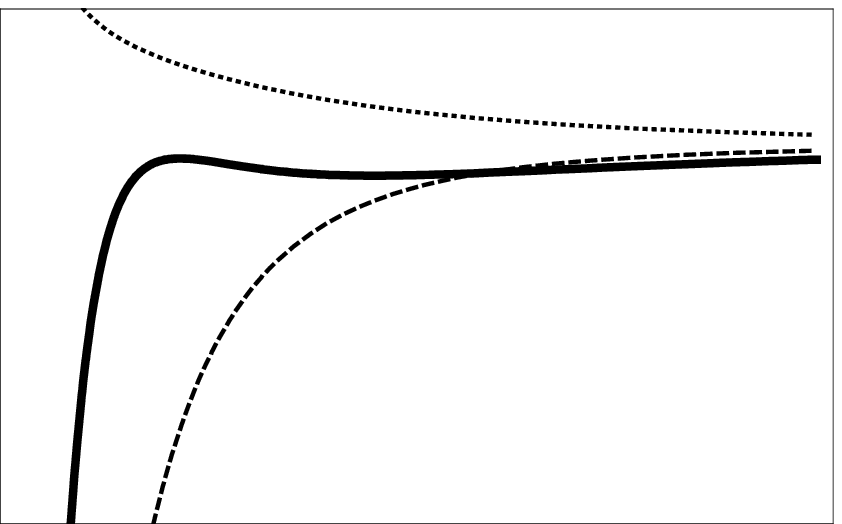,width = 10cm}
\caption{The back-reacted energy momentum tensor at weak coupling. The 
solid black line is the energy, the dashed line the radial tension, and
the dotted line the angular tension.}
\label{fig:weakcoup}}       % Give a unique label

Hopefully these lectures have given an insight into the complex and
fascinating topic of braneworld black holes. However the field develops 
over the next few years, there are sufficient puzzles and unanswered questions
to ensure that it will continue to be an active and exciting area.  

\acknowledgments
I would like to thank Lefteris Papantonopoulos for inviting me to such
a lovely school, and also my collaborators throughout the years 
but in particular Simon Creek, Yiota Kanti,
Bina Mistry, Simon Ross, Richard Whisker and Robin Zegers.
This work was partially supported by the EU FP6 Marie Curie Research 
\& Training Network "UniverseNet" (MRTN-CT-2006-035863)

\end{document}